\documentclass[fleqn,usenatbib]{mnras} 

\usepackage{newtxtext,newtxmath}

\usepackage[T1]{fontenc}
\usepackage{pifont}
\DeclareRobustCommand{\VAN}[3]{#2}
\let\VANthebibliography\thebibliography
\def\thebibliography{\DeclareRobustCommand{\VAN}[3]{##3}\VANthebibliography}
\usepackage{graphicx}	
\usepackage{amsmath}	
\usepackage[dvipsnames]{xcolor}
\usepackage{units}
\usepackage{bm}
\usepackage{threeparttable}
\usepackage{subcaption}

\title[FO Cepheids in the MCs]{\textbf{Multiwavelength study of observed and predicted pulsation properties of First overtone Cepheids in the Magellanic Clouds}}

\author[Kurbah et al.]{Kerdaris Kurbah$^{1}$\thanks{E-mail: kerdarisbahkur@gmail.com},
Shashi M. Kanbur$^{2}$\thanks{E-mail:shashi.kanbur@oswego.edu},
Sukanta Deb$^{1,3}$\thanks{E-mail:sukanta.deb@cottonuniversity.ac.in},
Anupam Bhardwaj$^{4}$,
Mami Deka$^{5}$,
Susmita Das$^{4,6,7}$,
\newauthor
Gautam Bhuyan$^{1}$\\
\\
$^{1}$Department of Physics, Cotton University, Guwahati 781001,
Assam, India \\
$^{2}$Department of Physics, State University of New York, Oswego, NY 23126, USA \\
$^{3}$Space and Astronomy Research Centre, Cotton University, Guwahati 781001, Assam, India \\
$^{4}$Inter-University Centre for Astronomy and Astrophysics (IUCAA), Post Bag 4, Ganeshkhind, Pune 411 007, India\\
$^{5}$INAF-Osservatorio astronomico di Capodimonte, Via Moiariello 16, I-80131 Napoli, Italy\\
$^{6}$Konkoly Observatory, Research Centre for Astronomy and Earth Sciences, HUN-REN, Konkoly-Thege Mikl\'os \'ut 15-17, H-1121, Budapest, Hungary\\
$^{7}$CSFK, MTA Centre of Excellence, Budapest, Konkoly Thege Miklós út 15-17., H-1121, Hungary}

\begin{document}
\label{firstpage}
\pagerange{\pageref{firstpage}--\pageref{lastpage}}
\maketitle
\begin{abstract}
We present a detailed analysis of the light curves and pulsation properties of First Overtone (FO) Cepheids in 
the Magellanic Clouds (MCs) obtained using observations and predictions from stellar pulsation models. 
Multiwavelength observational light curves were compiled from the literature (OGLE-IV, Gaia and VMC). We investigate the period-amplitude (PA), period-colour (PC), period-luminosity (PL), and 
amplitude-colour (AC) relations for FO Cepheids at multiwavelengths. We find that the PA distribution of FO Cepheids in the MCs modelled using a Gaussian Mixture Model shows that the SMC consists of higher amplitude stars than the LMC. \textbf{We find} multiple break-points in the PC/PL/AC relations for FO/FU Cepheids in the optical and near-infrared bands including the one near to $P=2.5$~d in the MCs using piecewise regression analysis and $F$-test statistics. Similarly, for the LMC FO Cepheids, we find a break-point in the PC/PL/AC relations near $P=0.58$~d. The slopes of the PC relations for LMC FO Cepheids are found to be shallow for $0.58<P(\rm d)<2.5$ but steeper for $P<0.58$~d and $P>2.5$~d. We complemented the observed relations using theoretical models for FO Cepheids with chemical compositions $Z= 0.008$ and $Z = 0.004$, appropriate for the LMC and SMC, respectively computed with MESA-RSP. Our results show that the pulsation properties of FO Cepheids in PC/PL/AC relations and colour-magnitude diagram are strongly correlated and their connections can provide stringent constraints for the theoretical pulsation models.
\end{abstract}

\begin{keywords}
	stars: variable: Cepheids-galaxies: Magellanic Clouds-methods: data analysis-methods: statistical 
\end{keywords}

\section{Introduction}
Classical Cepheids (CCs) are young intermediate-mass supergiant stars of F and G spectral types. CCs are 
abundant in both the Magellanic Clouds (MCs) and hence play a vital role in deciphering the various properties of these two 
galaxies such as the distance and geometry, including the evolutionary history of the MCs. The importance of 
Cepheids in stellar astrophysics dates back to the discovery of the existence of a PL relation, popularly known 
as Leavitt Law \citep{leav08,leav12}. Most Cepheids in the MCs are found to pulsate in the fundamental (FU, $\sim 52\%$) or first overtone (FO, $\sim 38\%$) mode, but some pulsate in the second overtone mode while others in double or triple modes \citep{sosz15, sosz17}. The MCs are two nearby satellite galaxies of the Milky Way (MW) with LMC and SMC being located at distances of approximately $50.0$ kpc and $62.5$ kpc, respectively \citep{piet19,grac20}. Several studies of resolved stellar populations in the literature have shown that the geometry of the LMC is roughly planar \citep[and reference therein]{niko04, deb14, more14, inno16, deb18, choi18, cusa21, ripe22, bhuy24} while the SMC shows a distorted shape being elongated up to 20 kpc towards the Milky Way \citep[and reference therein] {subr12,hasc12,more14,deb15,ripe16,ripe17,mura18, deb19,tatt21,zivi21, yanc21}. Similar studies have been carried out in tracing the structure of the MW \citep[and reference therein]{skow19, minn21, lema22,drim24}.  

Many studies have investigated the PL relations for FU Cepheids ranging from optical to near-infrared bands 
\citep[and reference therein]{bhar16a,sozy24}. In addition to the PL relations, the PC and AC relations of Cepheids and other variables such as RR Lyraes and $\delta$ Scutis have been extensively investigated in the literature \citep{simo93, kanb96,kanb03, sand04,sand09,kanb10, bhar16a, das20, deka22,kurb23}. The physics and the correlation
of the PC and PL relations stem from the period-luminosity-colour (PLC) relation \citep{mado91}. The changes in one relation are reflected on the other and vice versa \citep[and reference therein]{kanb04}.

The PA relation is not as straightforward as the other correlations between the 
observable physical properties of Cepheids. It depends on several factors such as period, metallicity, 
companionship, position in the instability strip (IS), and pulsation mode/energy. Based on empirical relations,
in the 1970s, \citet{mado76,butl76,butl78,fern70,fern90,coga80,sand71} have shown the dependence of Cepheids 
amplitude on the position in the IS (see \citet{bono00} for a brief explanation). Using canonical and non-
canonical models constructed from the mass-luminosity (ML) relations with and without core overshooting evolutionary models, respectively. \citet{bono00} found that for short-period ($P<10$~d) FU Cepheids, the models acquire maximum amplitude near the blue edge of the IS. However, the same study could not reach to a 
conclusive argument for FO Cepheids because of the smaller number of models. Furthermore, \citet{klag09,szab12} 
have demonstrated that with proper grouping, metal-deficient FO and short-period FU Cepheids in the Galaxy tend 
to pulsate with larger amplitudes. On the other hand, \citet{maja13} found that for long-period Cepheids 
($P>10$~d) in metal-poor galaxies, the metal-rich Cepheids display larger $V$-band amplitudes
than their metal-poor counterparts. A more detailed investigation with accurate models will shed more light on the 
understanding of the amplitude of Cepheids. 

FO Cepheids are of shorter periods ($\sim 0.2<P<9.43$~d) \citep{sosz15,piet21} as compared to the FU 
Cepheids ($\sim 0.7<P<208$~d), but comparatively brighter than the FU Cepheids in the overlapping period range 
between $\sim1.0<P<6$~d in the PL plane \citep{sosz15}. Because of their higher luminosities, numerous 
studies in the literature widely used FU Cepheids as distance calibrators in the first rung of the cosmic 
distance ladder \citep[and references therein]{ries22}. However, FO Cepheids have shorter periods and 
lower luminosities than the long period FU mode Cepheids which make them relatively fainter. Therefore, 
detecting these objects at larger distances becomes increasingly difficult.  Nevertheless, FO Cepheids combine with FU Cepheids,  play a significant role in tracing the structure of nearby host galaxies using optical and near-infrared wavelengths \citep{bono02, inno13, ripe16,inno16, bhar16a, deb18, deb19, 
ripe22,bhuy24}.

Earlier and recent studies of the PL relations for FO and FU Cepheids in the LMC and SMC have reported the 
presence of break-points at certain periods. For FO Cepheids, a break-point in the PL relations at $P=2.5$~d in optical and near-infrared photometric bands has been obtained for both LMC and SMC FO Cepheids in the study of 
\citet{bhar16,bhar16a}. \citet{subr15} reported a break-point in the PL relations for FO Cepheids at $\log 
{P}\sim0.229$~d and $\log{P}=0.4$~d in the SMC. Recently, \citet{ripe22} reported a break-point in the PL relation of the FO Cepheids in the LMC at $P=0.58$~d. Similarly, the break-point in the Cepheids PL relation for short-period FU Cepheids at $P=2.5$~d is well known in the literature \citep{ngeo10,tamm11,subr15,bhar16a}. The break-point at $P=2.5$~d is also found in the PC and AC relations for FU/FO Cepheids in the LMC and SMC \citep{bhar14}. \citet{refId0} found a break-point in the empirical instability strip (IS) at $P \sim 3$~d for both FU and FO Cepheids in the LMC which they attributed to the depopulation of second and third-crossing Classical Cepheids in the fainter part of the IS. In the present study, we carried out a detailed theoretical model calculations using MESA-RSP in multi-wavelengths (OGLE-IV, Gaia and VMC passbands) to confirm the presence of break-points in the PC/PL/AC relations for FO/FU Cepheids in the MCs obtained from observations as reported in the literature. 

Several efforts have been made to model pulsating variables such as Cepheids \citep{dorf91, yeck98, buch00, fior07, 
somm20, das20, das21, somm22, das24, deka24, marc24} and RR Lyraes \citep{feuc99a, feuc99, koll00, gabo23, gabo24} 
using non-linear hydrodynamical 1D models. From the theoretical perspective, a few studies have investigated the evolutionary and pulsating behaviour of FO Cepheids \citep{chio93,bono98,bono99a, feuc00, koll00, fior07,somm22}. These authors have shown that the input physics and the flux carriers in modelling the Cepheids FO models play a crucial role in the pulsation, luminosities, instability strip and the maximum periods of FO Cepheids for different chemical compositions. In addition, a study by \citet{anto93,kien99, feuc00,buch04} has shown that the $P_{4}/P_{1}=0.5$ resonance between the first and the fourth overtone located at $P\sim 4$ d plays a vital role in the light curve structure of the FO Cepheids.  Similarly, the resonance effect can also be seen in Hertzsprung progression in the case of FU Cepheids and beat pulsation for double mode Cepheids \citep{stob69,simo76,buch86, buch90,mosk92,marc00, buch04, smol10}. FO Cepheids thus provide additional and stringent constraints on theoretical models and the pulsation theory. However, the light curves of FO Cepheids have not been studied in details using \textsc{mesa-rsp}. In a recent paper, \citet{kurb23} (hereafter, Paper I) computed a grid of the FU Cepheids models using version r15140 of non-linear Radial Stellar Pulsation (RSP) within \textsc{mesa} \citep{paxt11, paxt13, paxt15, paxt18, paxt19}. In Paper I, the effect of using different convection sets on Fourier parameters and the PC, PL, AC, and PA relations of FU Cepheids in the MCs has been studied. These parameters and the relations are then compared with those obtained from observations. A comparison between the theoretical parameters/relations with the empirical ones provides useful information to constrain theories and models of stellar pulsation and evolution 
\citep{simo85,bono00, marc13, bhar17, bhar17a, das18, das20, deka22}. 

In continuation of Paper I, this study aims to investigate the PC/PL/AC relations, CMD and the Fourier 
parameters of the light curves of FO Cepheids obtained from observations and models. It is important to
note that this study constitutes about $\sim 1000$ non-linear full amplitude stable pulsation models; one of the largest sets of FO models computed using \textsc{mesa-rsp}. The comparison between the observations and the 
models in the FP/PC/PL/AC/CMD planes provides a rigorous way to constrain the pulsation models. In addition, 
investigating the break-points in these planes further enhances the understanding of the 
interconnection among these relations. 

The structure of the paper is organised as follows: the data and methodology are discussed in Section \ref{sec:data}. Section~\ref{sec:results} deals with the results and discussions related to the Fourier Parameters and various relations such as the PC/PL/AC relations including the CMD. The summary and the conclusion of the present study are provided in Section \ref{sec:conclusion}.

\section{Data and Methodology}
\label{sec:data}
\subsection{Observations}
\subsubsection{Optical Data}
\label{sec:optical_data}
The light curve of FU and FO Cepheids in the optical $(V,I)$ bands for both LMC and SMC are taken from 
the OGLE-IV catalogue \citep{sosz15, sosz17}. To avoid any uncertain characterization stars in the sample, 
we have discarded those stars listed in the ``remarks.text'' provided by OGLE-IV. There are $2477$ FU and $1780$
FO Cepheids in the LMC, whereas $2754$ FU and $1800$ FO Cepheids in the SMC in the OGLE-IV database. We have 
considered only those stars with $P<10$~d. For both FU and FO Cepheids, we choose complementary light curves in
both bands with more than 20 data points. This leaves us with 1459 and 1625 FO Cepheids in the LMC and the SMC, respectively. Similarly, this study aims to study the short period FU Cepheids with $P<10$ d. The number of short period FU Cepheids are $1931, 2321$  for LMC and SMC, respectively. 

Using the right ascension and declination ($\alpha, \delta$) values of Cepheids of the target galaxies available from the OGLE - IV catalogue, we use the CDS Xmatch \footnote{\url{http://cdsxmatch.u-strasbg.fr/}} to obtain the 
identification in the Gaia photometric bands: $G, G_{RP}$ ($G$: green photometer, $G_{RP}$; red photometer) in the Gaia DR3 survey \citep{gaia23}. Gaia DR3 is the third data release of the European Space Agency's (ESA) Gaia mission \citep{gaia16a}. Using the identification obtained from the CDS Xmatch, we retrieve the photometric light 
curves in these two bands. Cross-matching the data yield $1005$/$1312$ FO Cepheids as well as $1669$/$1637$ FU Cepheids in the LMC and SMC, respectively. Many of the observed light curves in the Gaia photometric bands are sparsely sampled. Therefore, we selected only those Gaia light curves having more than 20 data points.

\subsubsection{Cleaning of the Data}

\begin{figure*}
\begin{tabular}{cc}
\resizebox{1.0\linewidth}{!}{\includegraphics*{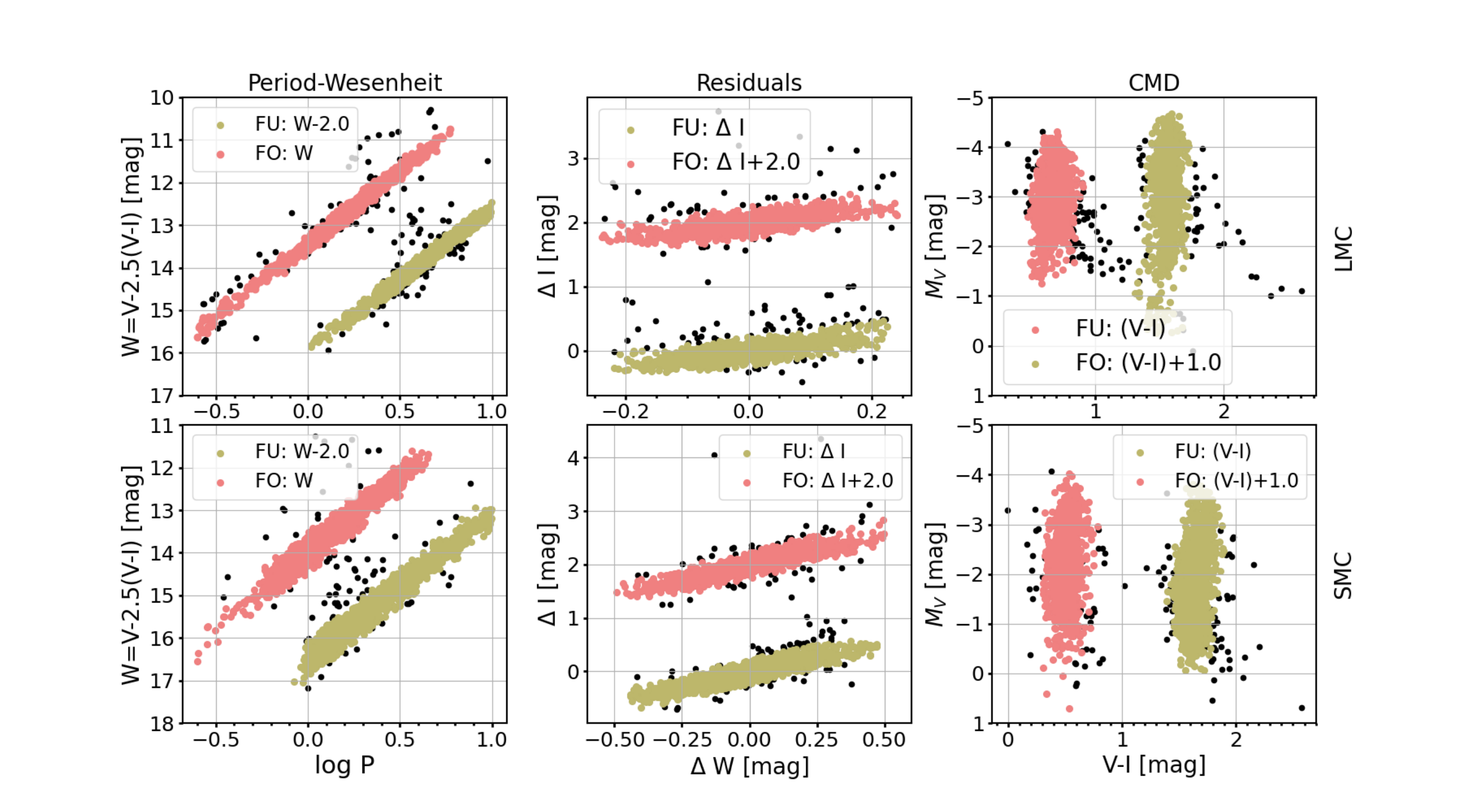}}& 
\end{tabular}
\caption{Left Panel: Period-Wesenheit relations for FU/FO Cepheids in the LMC and SMC. Middle Panel: Magnitude 
residuals for the I-band PL relations ($\Delta I_{0}$) and PW relations ($\Delta W$) for the same. Right Panel: CMD
diagram for the same. Orange and light green are the FU and FO Cepheids in the LMC. Black dots in all the plots 
are the discarded data points that deviate by more than 3 sigma. }
\label{fig:pw_res_cmd}
\end{figure*}

\begin{figure}
\begin{tabular}{cc}
\resizebox{1.0\linewidth}{!}{\includegraphics*{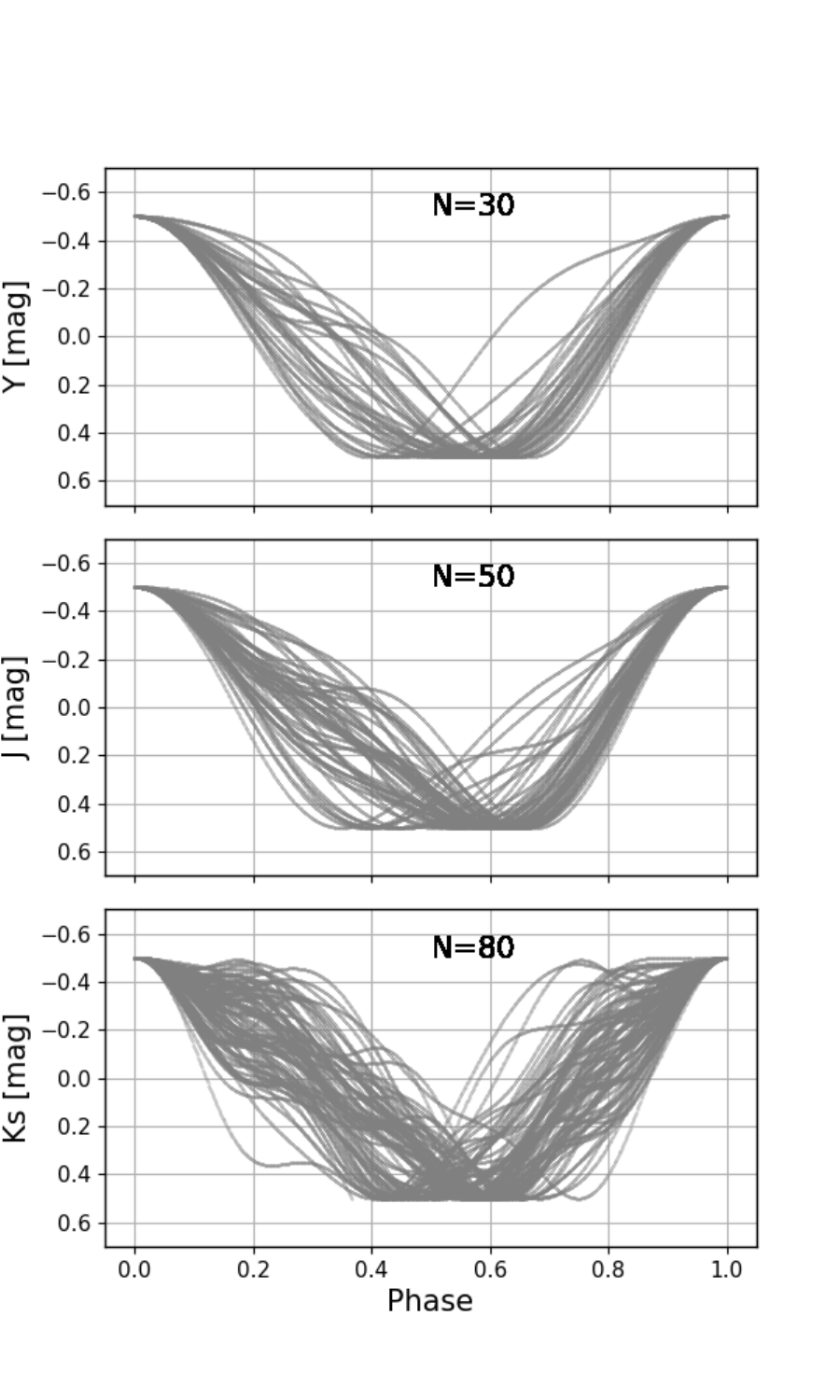}}& 
\end{tabular}
\caption{Light curve templates for LMC FO Cepheids in $Y, J, Ks$ bands. }
\label{fig:template_lmc}
\end{figure}

\begin{figure}
\begin{tabular}{cc}
\resizebox{1.0\linewidth}{!}{\includegraphics*{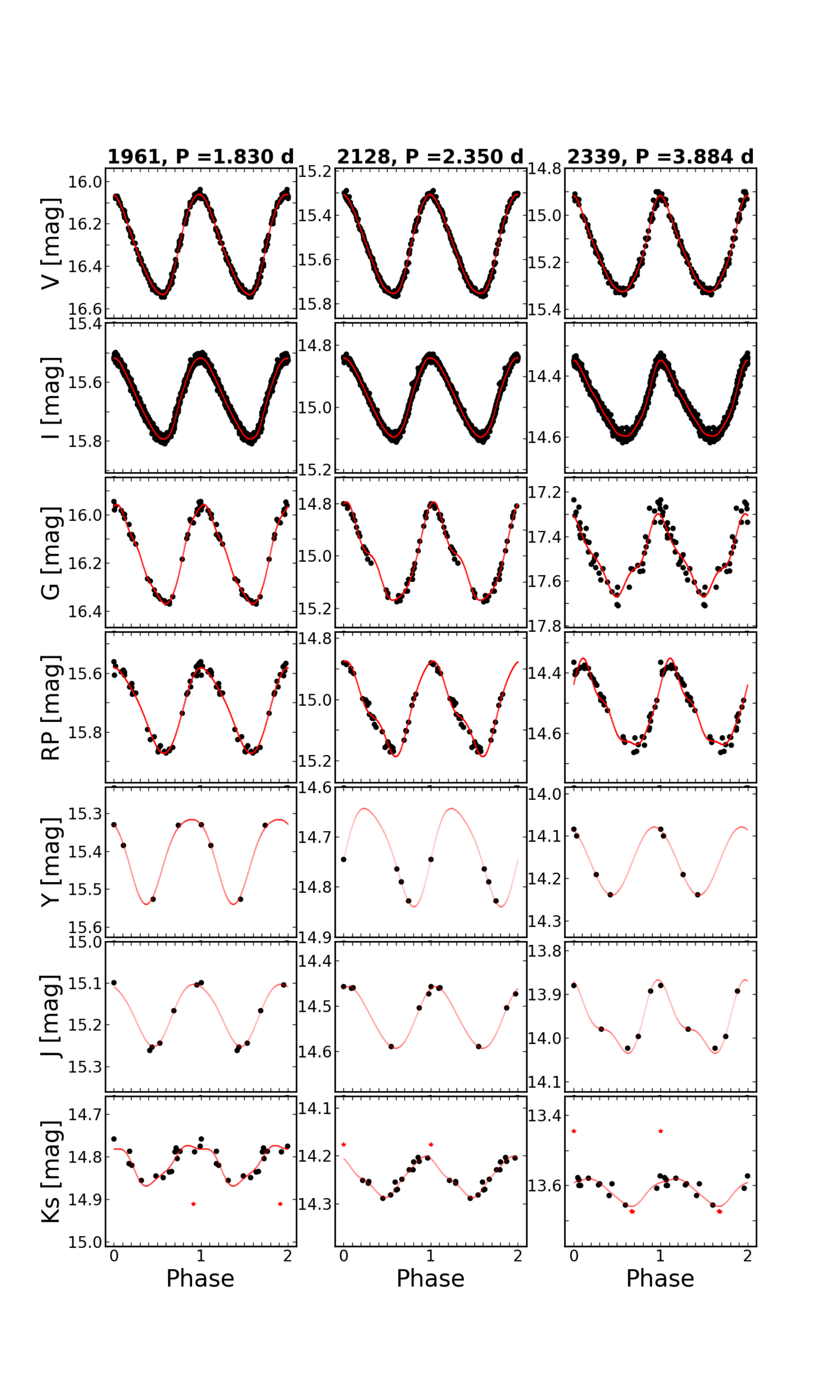}}& 
\end{tabular}
\caption{Fitted light curves for three LMC Cepheids in multiple bands ($VIGG_{RP}YJKs$). The star IDs consisting
of four numbers are based on the OGLE-IV catalogue. Black dots in all the plots represent the observed data and the red star symbol represents the outliers. The solid red line represents the template fit for near-infrared bands ($YJKs$) and Fourier fit for optical bands ($VIGG_{RP}$).  }
\label{fig:lc_yjks}
\end{figure}

The outliers in the raw light curves for all the photometric bands are removed using the condition 
\citep{leys13, deka22}:
\begin{align}
\frac{|m-\textrm{Median}(m)|}{MAD(m)} \ge 3.0.
\end{align}
Here m is the observed mean magnitude and MAD represents the median absolute deviation.

The light curves in the OGLE-IV and Gaia photometric bands are fitted using a Fourier decomposition cosine series 
\citep{deb09}:
\begin{align}
m(t)= & A_{0} + \sum_{i=1}^{N} A_{i}\cos\left(2\pi i\Phi + \phi_{i}\right).
\end{align}
We obtain the mean magnitudes, amplitudes and the Fourier phase/amplitude parameters from the fitted light 
curves along with their uncertainties. Here $A_{i}$ and $\phi_{i}$~denote the Fourier coefficients, where 
$i=1,2,3,...N$. The optimal value of $N$ is obtained using the Baart's Criteria \citep{baar82,pete86}. For the
near-infrared bands, the light curves are phase folded using the the values of period and epoch available from the OGLE-IV catalogue \citep{sosz15}. 

For the LMC and SMC FU/FO Cepheids, we determine the reddening free Wesenheit index $W_{\textsc{vi}}=I-2.55(V-I)$ \citep{mado91}. The period-Wesenheit (PW) relations are plotted in the left panel of Fig.~\ref{fig:pw_res_cmd}. We apply $3\sigma$ clipping to remove the outliers. Residuals obtained from the PW relations ($\Delta W$) and the PL relations in $I$-band ($\Delta I$) are shown in the middle panel of Fig.~\ref{fig:pw_res_cmd}. Stars that deviate above $3\sigma$ from the relations between the magnitude residuals of the PL relation in I-band and the corresponding residuals of the PW relations are rejected. These large deviations may be due to uncertainties related with the apparent magnitudes and reddening values for each star \citep{mado17}.
The CMD for the LMC/SMC Cepheids is shown in the right panel of Fig.~\ref{fig:pw_res_cmd}. The 
black dots in all the plots shown in Fig.~\ref{fig:pw_res_cmd} denote the rejected data points. 
After the cleaning process, the summary of the LMC and SMC FO/FU light curves in $VIGG_{RP}$ photometric bands are provided in Table~\ref{tab:number_obs}.

\subsubsection{Near-infrared Data}
The light curves in the near-infrared bands ($YJKs$) for both LMC and SMC are obtained from the VMC (The $VISTA$ near-
infrared $YJKs$ survey of the Magellanic System) database. The VMC survey is a near-infrared survey that covers
an area of $\rm \sim180~deg^{2}$ of the MCs, consisting of $\rm 116~deg^{2}$ of the LMC and $\rm 45~deg^{2}$ of
the SMC \citep{cion11, ripe16, ripe22}. Retrieving the Cepheids in the VMC survey using the OGLE-IV 
identifications yields $1354$ FO, $1807$ FU stars in the LMC and $1424$ FO, $2213$ FU stars in the SMC. The number of stars in $YJKs$ photometric bands, after we cross-matched with the clean sample obtained in the previous section, is summarised in Table~\ref{tab:number_obs}.

Following the procedure as described in \citet{ripe22}, we determine the best-fit templates and model the 
complimentary light curves in the respective $YJKs$ bands. We visually select the best Fourier-fitted phased light curves for each photometric band to generate the template light curves. The chosen light curves are of various shapes and include a wide range of periods. For FO Cepheids, the selected templates consist of $30, 50, 80$ model light curves in $YJKs$ bands for the LMC and $8$ model light curves in each band for the SMC. Also, the selected templates for FU Cepheids in each band contain $38, 38,$ and $132$ model light curves for the LMC. Similarly, the model light curves in the case of SMC are $15, 27$ and $73$ in $Y, J, Ks$ bands, respectively. Fig.~\ref{fig:template_lmc} displays the light curve templates for FO Cepheids in the LMC in $YJKs$ bands. Each model light curve is subtracted from the intensity averaged mean magnitude. The subtracted light curves are normalized so that the mean magnitudes are equal to $0.0$ and the amplitudes are $1$. Each template is fitted to the light curves by shifting the magnitudes and phase to obtain the mean magnitudes and amplitudes based on the minimum value of $\chi^{2}$. Fig.~\ref{fig:lc_yjks} displays three light curves of LMC FO Cepheids in multiple bands ($VIGG_{RP}YJKs$). Black dots in all the plots represent the observed data and the red star symbol represents the outliers. The red line represents the template/Fourier fits for near-infrared and optical bands, respectively.  

\begin{table}
\begin{center}
\caption{Number of LMC and SMC samples used in the present study. N is the number of stars.}
\begin{tabular}{c c c c c c c c c} \\ \hline  \hline
Bands & Galaxy & Class &    N & Source\\ \hline
 $V$    & LMC   & FU    & $1771$    & \citet{sosz17}\\
        &       & FO    & $1355$    & ,, \\
        & SMC   & FU    & $2196$    & ,,\\
        &       & FO    & $1532$    & ,, \\
 $I$    & LMC   & FU    & $1771$    & ,,\\
        &       & FO    & $1355$    & ,, \\ 
        & SMC   & FU    & $2196$    & ,,\\
        &       & FO    & $1532$    & ,, \\
 $G$    & LMC   & FU    & $1548$    & \citet{gaia16}\\
        &       & FO    & $795$     & ,, \\
        & SMC   & FU    & $1634$    & ,,\\
        &       & FO    & $1182$    & ,,\\
 $G_{RP}$   & LMC   & FU    & $1616$    & ,,\\
            &       & FO    & $988$     & ,, \\
            & SMC   & FU    & $1633$    & ,,\\
            &       & FO    & $1162$    & \\
 $Y$    &   LMC & FU    & $1734$    & \citet{ripe22}\\
        &       & FO    & $1291$    & ,, \\
        &   SMC & FU    & $1895$    & \citet{ripe16}\\
        &       & FO    & $1425$    & ,, \\
 $J$    &   LMC & FU    & $1482$    & \citet{ripe22}\\
        &       & FO    & $1291$    & ,, \\
        &   SMC & FU    & $1895$    & \citet{ripe16}\\
        &       & FO    & $1425$    & ,, \\
 $Ks$   &   LMC & FU    & $1734$    & \citet{ripe22}\\
        &       & FO    & $1291$    & ,, \\    
        &   SMC & FU    & $1895$    & \citet{ripe16}\\ 
        &       & FO    & $1425$    & ,,  \\ \hline \hline
\end{tabular}
\label{tab:number_obs}
\end{center}
\end{table}

\subsubsection{Extinction Correction}
We use the reddening map of \citet{skow21} to obtain the extinction-corrected mean magnitude in all the selected photometric bands  ($V, I, G, G_{RP}, Y, J, Ks$) considered in this study. The right ascension ($\alpha$) and declination ($\beta$) of both FO/FU Cepheids in the MCs as available from the OGLE-IV database are used to obtain the corresponding reddening values from the reddening map. The central wavelengths of the seven photometric bands are $\lambda_{V, I, G, G_{RP}, Y,J, Ks}=(0.556, 0.810, 0.673, 0.797, 1.020, 1.254, 2.149)~\mu$m \citep{bona10,jord10,cion11,wang19,iwan22}. Applying the \citet{card89} reddening law with a fixed value of $R_{V}=3.23$, the corresponding values of total-to-selective absorption are obtained as $R_{V, I, G, G_{RP}, Y, J, Ks}=(3.23, 2.05, 2.62, 2.03, 1.28, 0.92, 0.39)$. The correction is made using the extinction law $A_{\lambda} = R_{\lambda} E(B-V)$, where the $E(B-V)$ values are obtained by converting $E(V-I)$ obtained from the reddening map using the relation, $E(V-I)=1.26 E(B-V)$ following the standard Cardelli law.

\subsection{Theoretical Models}
\label{sec:fo_models}

\begin{figure*}
\begin{tabular}{c}
\resizebox{1.0\linewidth}{!}{\includegraphics*{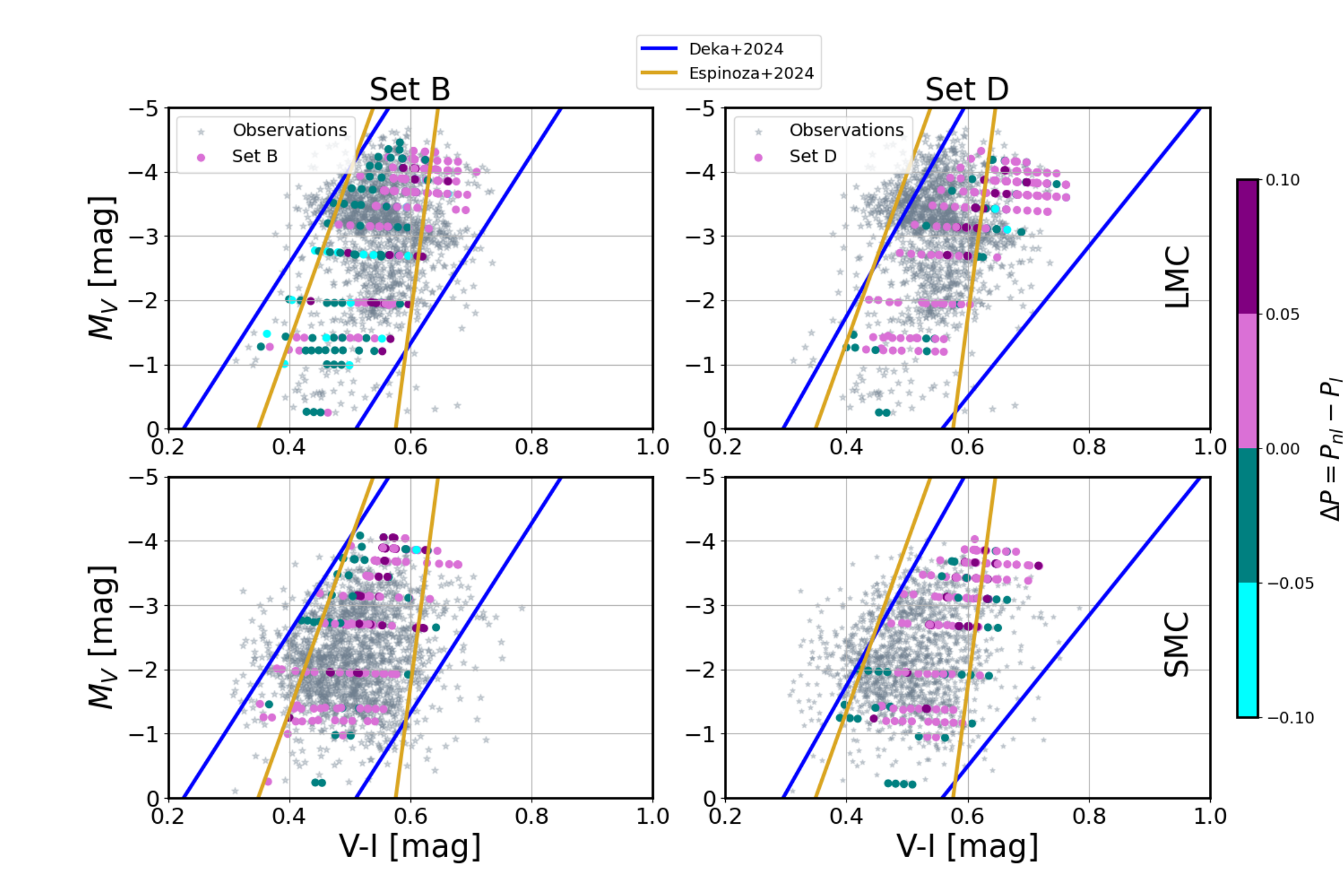}}
\end{tabular}
\caption{CMD for LMC and SMC FO Cepheids obtained using the observations and models. The IS edges are taken from \citet{deka24} and \citet{espi24} . }
\label{fig:cmd_lmcsmc_fufo_obsmodels}
\end{figure*}

We have computed the Cepheids models using an open-source code, \textsc{mesa-rsp} (version r15140) 
\citep{paxt11, paxt13,paxt15,paxt18,paxt19}. The theory of convection and the stellar pulsation treatment as 
implemented in \textsc{mesa-rsp} are based on the works of \citet{kuhf86} and \citet{smol08}. For this study we have used the four sets of convection parameters labelled as convection sets A, B, C, and D as provided in \citet{paxt19} to explore the effects of these convection sets in modelling the pulsational properties of FO Cepheids. These four convection sets are defined based on the different combinations of the eight free parameters ($\alpha, \alpha_{m}, \alpha_{p}, \alpha_{t}, \alpha_{s}, \alpha_{c}, \alpha_{d}$ and $\gamma_{r}$). Set A corresponds to the simplest
convection model without turbulent pressure, turbulent flux, and radiative cooling ($\alpha_{t}=0.0, \alpha_{t}
=0.0, \gamma_{r}=0.0$ ), set B adds radiative cooling without turbulent pressure and turbulent flux ($\alpha_{t}
=0.0, \alpha_{t}=0.0, \gamma_{r}=1.0$), set C adds turbulent pressure and turbulent flux without radiative 
cooling ($\alpha_{t}=1.0, \alpha_{t}=0.01, \gamma_{r}=0.0$) and Set D includes these effects simultaneously 
($\alpha_{t}=1.0, \alpha_{t}=0.01, \gamma_{r}=1.0$). The values of $\alpha_{m}$ for the convection sets A, B,
C, and D are $0.25, 0.50, 0.40$, and $0.70$, respectively. $\alpha_{m}$ was adjusted so that the models with 
different sets have similar amplitudes. The other four parameters ($\alpha, \alpha_{s}, \alpha_{c}, 
\alpha_{d}$) are the same for all the sets. In this study, we have adopted these four convection sets as 
outlined in \citet{paxt19}. However, as highlighted in \citet{paxt19}, these parameter values are merely useful starting choices and calibration of the convection parameters with multiple observational constraints is indeed crucial. The inlist ($\rm inlist\_rsp\_common\_pulsation$) used in computing the pulsation models is available at \url{https://github.com/Kerdaris/FO_MC_CMD}.

The chemical compositions adopted for the LMC and SMC are $Z = 0.008, X = 0.742$ and $Z = 0.004, X = 0.746$, 
respectively \citep{bono98}. The stellar masses for FO Cepheids are chosen in the range of 3.0 $M_{\odot}$ to 
6.8 $M_{\odot}$ which is an ideal mass range to obtain the Cepheids in the first overtone mode \citep{bono99}. 
The effective temperature is between $4900-6550$ K in step of $50$. For a particular mass, the luminosity 
of the model is calculated using the rotation-averaged ML relation of \citet{ande14}. The upper and lower limit
of the luminosity calculated are set by increasing/decreasing the luminosity level by $\Delta \rm log L/
L_{\odot}= 0.25$~dex. A linear non-adiabatic (LNA) radial pulsation analysis has been carried out to obtain the radial pulsation mode, growth rates of the radial pulsation mode, and the linear pulsation period. For both LMC and SMC, the models that exhibit positive growth rates in the FO and FU modes are selected from the linear calculation (LNA analysis). However, note that the final mode of pulsation can only be determined from non-linear computations. Also, using the linear period obtained from the linear calculation, we made a period cut for the LMC/SMC models at $P=6.0$~d and $P=4.6$~d, respectively. The period cut is made in accordance with the longest period FO 
Cepheids observed by the OGLE-IV survey in the LMC ($P=6.0$~d) and SMC ($P=4.5$) d \citep{sosz15}. The models selected from the linear calculations serve as input for the non-linear calculations. 

The non-linear integration is computed over a varying number of pulsation cycles till the model reaches the full amplitude stable pulsation (FASP) state and attains a single periodicity in the FO mode. The FASP state of the model is satisfactory provided the amplitude of radius variation ($\Delta R$), the pulsation period $P$ should not vary by more than $\sim 0.001$ over the last $\sim 1000$ cycles of the total integration computed ($\sim 4000$ pulsation cycles). The fractional growth of the kinetic energy per pulsation period ($\Gamma$) must approach zero once the FASP is reach. The condition of FASP is significant to ensure that the models has attained a stable state and not just in a transient phase of mode switching from fundamental to first overtone and vice-versa. From the FASP condition, some models were found to converge to single periodicity in the FU mode, and were subsequently excluded. Only the models that converge in the FO mode at FASP are chosen for further analysis. From the chosen models, the difference between the linear period ($P_{l}$) obtained from the LNA analysis and the non-linear period ($P_{nl}$) obtained from the FASP model, i.e $\Delta P = P_{nl}-P_{l}$ were calculated. The CMD obtained using the models and the observations are shown in Fig.~\ref{fig:cmd_lmcsmc_fufo_obsmodels}, the colourbar in each plot represents $\Delta P$. The edges of the IS in the CMD plane shown were adopted from \citet{espi24} and \citet{deka24}. The individual mean difference between the linear and non-linear periods in our models is $\sim 0.02$ d. Fig.~\ref{fig:HR_lmcsmc_bd} in Appendix \ref{sec:hrd} depicts the HR diagram (HRD) of the LMC/SMC models using convection sets B and D used in this study. The edges of the IS in the HRD shown were adopted from \citet{somm22}, \citet{espi24} and \citet{deka24}.
 
Table~\ref{tab:models_all} summarised the number of non-linear models that converge in the FO mode and reach the FASP. For the exact same set of $MLTXZ$ input parameters, we find a greater percentage of Cepheids models satisfying the condition of FASP and pulsating in the FO mode after non-linear integrations computed using sets B and D as opposed to sets A and C, for both LMC and SMC. The bolometric light curve is one of the default outputs of the non-linear computation of \textsc{mesa-rsp}. The default pre-computed bolometric correction tables available in \textsc{mesa} convert the bolometric magnitudes into the absolute magnitudes in $UBVRI$ photometric bands \citep{bess90} and $UBVRIJHKLL'M$ photometric bands \citep{leje98}. Similarly, we additionally included the bolometric correction table from MIST \footnote{https://waps.cfa.harvard.edu/MIST/index.html} (mesa Isochrones \& Stellar Tracks) packaged model grid \citep{choi16, dott16} in \textsc{mesa} to transform the bolometric magnitudes into absolute magnitudes in the Gaia-passbands ($G, G_{RP}$). The bolometric correction table from MIST is computed using 1D atmosphere models based on \textsc{ATLAS12/SYNTHE}. The table is defined as a function of $\rm log g$, $T_{\rm eff}$ and metallicity [M/H] \citep{kuru70,kuru93}. 

It is important to note that the models computed in this study do not encompass the entire instability strip (IS) where pulsations in the FO Cepheids are theoretically possible. Therefore, the models generated may not efficiently fill the parameter space of the observational data. We emphasize here that the theoretical PC/PL/AC relations of the FO Cepheids obtained from the present grid of models used in this study are preliminary investigations focused on FO Cepheids models. A more refined model grid addressing these shortcomings is a subject of future study.

The Cepheids FU models are taken from \citet{kurb23} where the theoretical light curves were available only in $V, I$ passbands with $P>2.5$~d. For $P<2.5$~d, we additionally computed the non-linear integration for Cepheids FU models using convection sets B and D. The choice of using only convection sets B and D was made because the analysis for FO Cepheid models were also done based on these two sets. Similar to the FO models, we check the condition of FASP state for the models computed. Models that converge to single periodicity in the FU mode are selected for the analysis. Table~\ref{tab:models_fu} summarised the number of non-linear FU models with $P<10$~d for both LMC and SMC.

\begin{table}
\begin{center}
\caption{Number of full amplitude stable pulsation models.}
\scalebox{0.88}{
\begin{tabular}{c c c c c c c c} \\ \hline  \hline
    &       & Set A & Set B & Set C & Set D & Total\\ \hline
Number of FASP models & LMC & $75$  & 278 & $19$ & 271   & 643 \\
Number of FASP models & SMC & $93$  & 224 & $27$ & 233   & 577 \\ \hline \hline
\end{tabular}}
\label{tab:models_all}
\end{center}
\end{table}

\begin{table}
\begin{center}
\caption{A summary of LMC and SMC FU/FO Cepheids models used in the analysis.}
\scalebox{0.86}{
\begin{tabular}{c c c c c c c c} \\ \hline  \hline
Type   & Sets   &  Pmax & Pmin & N  & $\rm M/M{\sun}$ & $\rm L/L{\sun}$ & $\rm T_{eff}$ (K) \\ \hline
\multicolumn{7}{c}{LMC} \\ \hline
FU  &  B  & $9.964$ & $0.648$ & $197$ & $3.0-6.8$ & $250-5500$ & $5350-6500$\\
    &  D  & $9.943$ & $0.724$ & $154$ & $3.0-6.8$ & $250-4500$ & $4700-6250$ \\
FO  &  B  & $5.876$ & $0.721$ & $278$ & $3.0-6.8$ & $300-4500$ & $5800-6500$\\
    &  D  & $5.983$ & $0.490$ & $271$ & $3.0-6.8$ & $250-4500$ & $5550-6450$ \\\hline
 \multicolumn{7}{c}{SMC} \\ \hline
FU  & B  & $9.897$ & $0.621$ & $205$ & $3.0-6.8$ & $250-5500$ & $4900-6600$\\
    & D  & $9.932$ & $0.495$ & $192$ & $3.0-6.8$ & $250-5000$ & $4900-6650$\\
FO  & B  & $4.592$ & $0.620$ & $224$ & $3.0-6.8$ & $250-3500$ & $5800-6600$\\
    & D  & $4.597$ & $0.495$ & $233$ & $3.0-6.8$ & $250-3500$ & $5700-6600$\\\hline \hline
\end{tabular}}
\label{tab:models_fu}
\end{center}
\end{table}

\section{Results and Discussion}
\label{sec:results}
In this section, we investigate the FP, PA, PC, PL, AC relations and the CMD for the FO and short-period FU ($P<10$~d) Cepheids in both LMC and SMC obtained from models and observations. The investigation on PC/PL/AC relations in the context of stellar photosphere and hydrogen ionization front interactions have been primarily carried out by \citet{simo93, kanb96, kanb03, kanb04,sand04, kanb05, kanb06, kanb07, kanb10, bhar14, bhar16a, das20, das21, deka22}. The mean light PC/PL/AC relations have been the subject of various studies for applications ranging from distance measurements to reddening maps \citep{sand93, ripe16, breu22, reis22}. In a series of papers, Sandage and Tammann investigated the PC relations in MW, LMC, and SMC, to understand the universality of Cepheids PL relation at mean light \citep{sand03, sand04, sand09}. Such studies provide a deeper understanding about the pulsation physics and the behaviour of the aforementioned relations at mean light. Regarding the break-points obtained in the literature, \citet{ripe22} reported a break-point in the PL relations for the LMC FO Cepheids at $P= 0.58$~d in the near-infrared bands at mean light. On the other hand, using OGLE-III data, \citet{bhar14, bhar16, bhar16a} have shown that the break-point in the PC/PL/AC relations for FO Cepheids occurs at $P=2.5$~d in both LMC and SMC using maximum and minimum light.

We investigate the presence of break-points in the empirical PC and PL relations for multi-passbands using piecewise regression analysis also known as segmented regression. This statistical technique is particularly effective in modeling and analyzing relationships that change across different intervals of an independent variable \citep[and references therein]{bai97, Pilg21}. In order to determine the optimal number of break-points,  we use the ModelSelection routine in the piecewise regression to ensure more reliable predictions. This is essential for the algorithm to converge successfully with the identification of the break-points within the data. It has been observed that the optimal number of break-points as well as their locations vary across different photometric bands. For an example, we have found breakpoints at $P=2.576$~d, $P=2.471$~d, $P=2.570$~d, $P=2.624$~d, $P=2.750$~d,  $P=2.630$~d, $P=2.618$~d  for $V, I,G, G_{RP}, Y, J, Ks$ bands, respectively using FO Cephieds in the SMC. Likewise, for the LMC FO Cepheids, the locations of the break-points are slightly different from each other across different bands. Furthermore, although the number of break-points for each band remains the same for the SMC, the same is not true for the LMC FO Cepheids. These imply that the locations of the break-points are not exactly the same and are slightly shifted from each other across different bands. This is in contrast to the reported value of a single break-point across different passbands as given in the literature. For example, a single break-point value of $P=2.5$~d has been reported in the literature for different optical and infrared bands \citep{ngeo10, tamm11, bhar16a}

Following the same procedure as those applied for the observed data, we have also attempted to determine the break-points from the theoretical data using piecewise regression. We could find the break-point values near $P=2.5$~d which differ from each other by a small amount across all the bands. Since the number of data points generated from the theoretical modelling is fewer in number and the data points are sparsely distributed, it is not possible for the piecewise regression to find all the break-points as those obtained using the observed data. Therefore, the theoretical PC/PL/AC relations of LMC and SMC FO/FU Cepheids were examined with a break-point determined near $P=2.5$~d. Also, we have left out the theoretical PC/PL/AC relations for FO models with $P<0.58$~d because of the smaller number of models below this period in comparison with the observed ones. Although the relations in the FP/PC/PL and AC planes are plotted as a function of period on a logarithmic scale, it is important to note that the discussion of the results obtained in these planes are given in days.

The significance of the break-points at multiple locations obtained using piecewise regression has been tested using a statistical $F$- test. The statistical $F$- test is described as follows: The null hypothesis is that a single regression line is consistent with the data, while the alternative hypothesis is that two regression lines are consistent with the data. The critical value of the $F$ statistic at $95\%$ confidence level is defined as $F_{c}=F_{2,n-4}$. The null hypothesis is rejected if the value of the $F$ statistic is greater than the critical value or the probability of the $F$ statistics $p(F)<0.05$ \citep{kanb04, bhar14}. 

We use the standard $t$-test to check the equivalence of theoretical PC/AC/PL slopes obtained using convection sets B, D and the observations. The detailed description of the $t$-test may be found in \citet{ngeo15}. The $T$ statistics in the $t$-test are mainly used to compare the slopes of two linear regressions.
If $\hat{W}$ represents the slope and Var($\hat{W}$), the variance of linear regression; the null hypothesis of 
the $t$-test is that the slope of a sample size $m$ ($\hat{W_{m}}$) is equivalent to the slope of a sample $n$ 
($W_{n}$). The $t$ statistics is defined as:
\begin{equation}
T=\frac{\hat{W}_{n}-\hat{W}_{m}}{\sqrt{\rm {Var} (\hat{W}_{n})-\rm {Var}(\hat{W}_{m})}}.
\end{equation}
The critical value under two-tailed $t$ distribution at $95\%$ confidence level is $t_{\alpha/2,\nu}$,
where $\alpha=0.05$. The degrees of freedom of $t$-test is $\nu=n+m-4$. The null hypothesis (equivalent slopes) is 
rejected if the probability of the observed value of the $T$ statistics is $p(t)<0.05$.

\subsection{Empirical PA relation}
\begin{figure*}
\begin{tabular}{cc}
\resizebox{1.0\linewidth}{!}{\includegraphics*{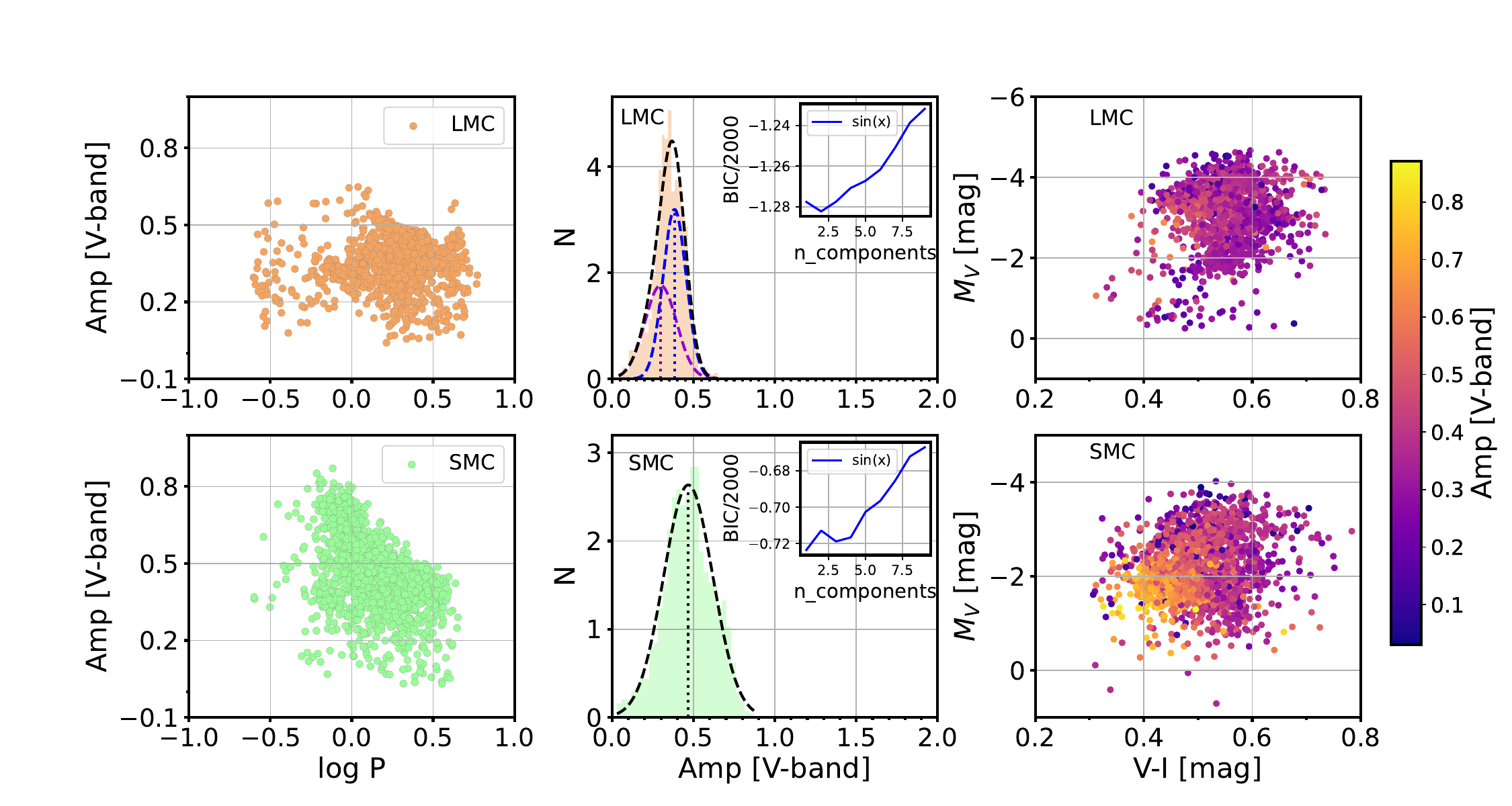}}& 
\end{tabular}
\caption{\textit{Left Panel}: PA distribution for the LMC (upper) and SMC (lower) FO Cepheids in $V$-band. 
\textit{Middle Panel}: Histogram of PA distribution is modelled using a Gaussian Mixture Model. The mean of the 
Gaussian fits for the FO Cepheids in the LMC/SMC in $V$-band are centered at ($0.298$, $0.383$) mag and 
($0.467$) mag, respectively. \textit{Right Panel}: Colour Magnitude Diagram with $V$-band
amplitude in the colour bar. It is evident from the CMD that there are more high amplitude stars in the SMC 
around a mean of $V-I=0.467$ mag.}
\label{fig:amp_cmd_dis_lmcsmc_fo}
\end{figure*}

Fig.~\ref{fig:amp_cmd_dis_lmcsmc_fo} represents the PA relation for FO Cepheids in the LMC (upper panel) and 
SMC (lower panel) for $V$-band. It can be seen from the distribution in the PA relation that the short period FO 
Cepheids (<$2.5$~d) in the SMC consist of high-amplitude stars as compared to those of the LMC within the same 
period range. Before we perform the Gaussian fit to the Period-Amplitude (PA) distribution, we tested the GMM by varying the number of components ($k$) from $1$ to $10$ using the Bayesian information criterion (BIC). The BIC values for each of the component are calculated. The number of components ($k$) of the GMM with the minimum BIC are chosen to fit the distribution. It is found that the PA relations for LMC and SMC FO Cepheids are best ﬁtted with $k = 2$  and $k = 1$, respectively corresponding to the lowest BIC value. The GMM fit corresponding to the $k$ values for both LMC and SMC are shown in the middle panel of Fig.~\ref{fig:amp_cmd_dis_lmcsmc_fo}. The amplitude distribution of FO Cepheids in the LMC and SMC modelled using a Gaussian mixture model are found to be centered at ($0.298$, $0.383$) mag and $0.467$ mag, respectively for $V$-band. It is evident from the middle panel of Fig.~\ref{fig:amp_cmd_dis_lmcsmc_fo} that the PA distribution indicates that the LMC consist of two population of high and low amplitude FO Cepheids. The mean obtained from the Gaussian fits from both the LMC and SMC suggest that the SMC contains higher amplitudes Cepheids as compared to the LMC counterpart. This result is consistent with the findings of \citet{klag09,szab12,deka22} that the metal-poor stars display larger amplitudes than their metal-rich counterparts. 

The right panel of Fig.~\ref{fig:amp_cmd_dis_lmcsmc_fo} displays the CMD of the FO Cepheids for both LMC and SMC. 
It can be seen from the plot that the SMC FO Cepheids with higher amplitudes have lower ($V-I$) colour values, thus 
are bluer and blue Cepheids are known to have larger amplitudes. Moreover, the high amplitude stars in the SMC are found around a mean of $V-I=0.467$ mag. This is consistent with the study of \citet[and
reference therein]{sand04, turn06} where it was found that the photometric amplitudes of Cepheids depend on 
their positions in the IS: larger near the blue side of the IS and the smaller one towards the red edge of the IS. 

\subsection{Comparison of the observed and theoretical light curves}
\begin{figure}
\begin{tabular}{cc}
\resizebox{1.0\linewidth}{!}{\includegraphics*{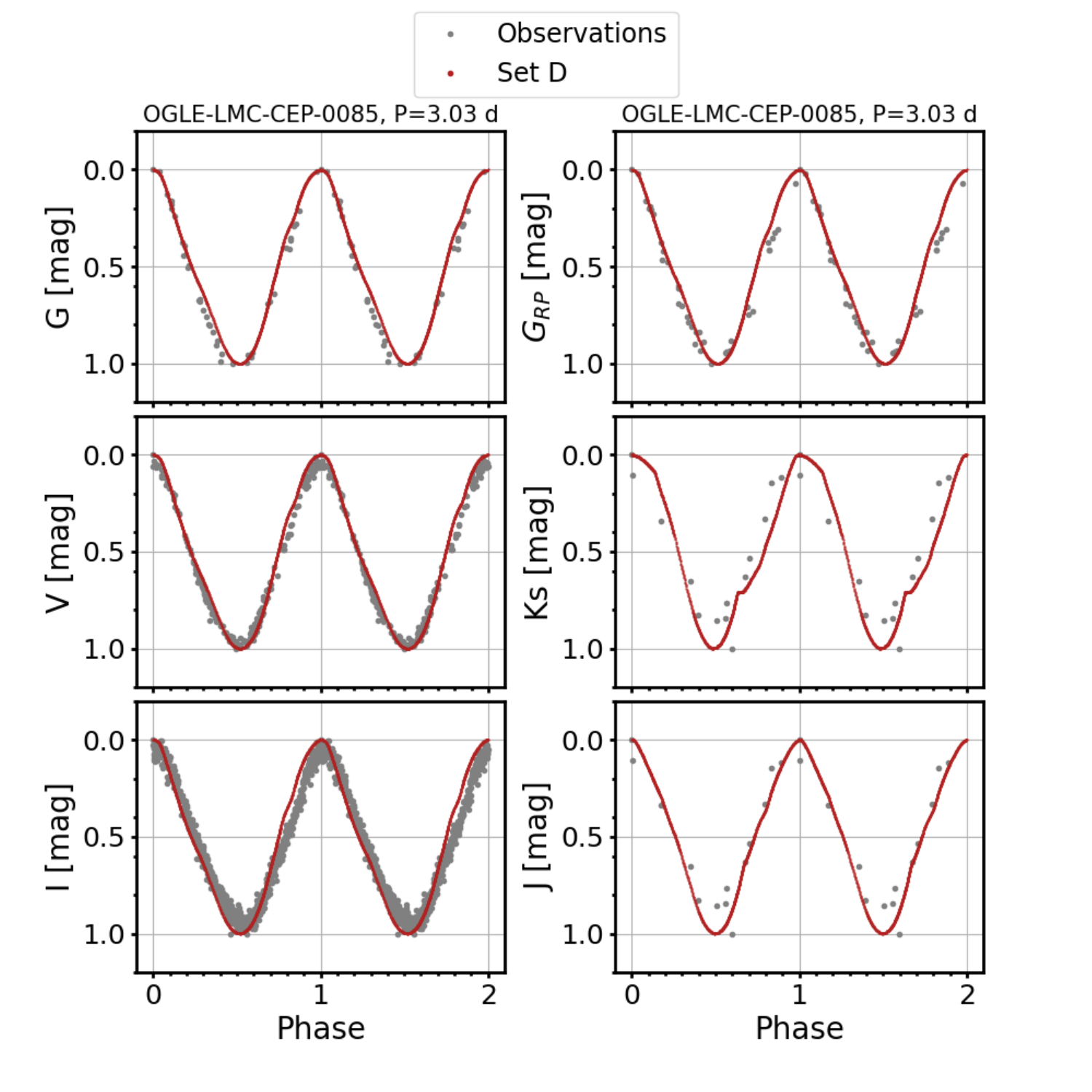}}& 
\end{tabular}
\caption{Normalized light curves using Set D (maroon colour) and observations (grey dots) in multiple bands ($VIGG_{RP}
JKs$). The input parameters of the model, $Z=0.008, X=0.742, M/M_{\odot}=5.0, L/L_{\odot}=3400,T_{eff}=5420$.} 
\label{fig:light_curves}
\end{figure}

A comparison between the observed and the theoretical light curves in multi-photometric bands is shown in 
Fig.~\ref{fig:light_curves} with the observed light curve ID: OGLE-LMC-CEP-0085 and theoretical light curve 
generated with $M=5M\odot, L=3400L\odot, T_{eff}=5420 K$ for the LMC composition. The pre-processed table from 
\citet{leje98} provided by \textsc{mesa-rsp} does not contain $Y$ passband. The light curves obtained 
from the model computed using set D are shown for $VIGG_{RP}JKs$ passbands. The comparison has been made based 
on the similarity of periods. The periods of the observed and theoretical light curves are taken to be 
$P=3.03$~d and $P=3.01$~d, respectively. It is evident from Fig.~\ref{fig:light_curves} that the light curves 
obtained from the model are found to be consistent with the observed ones in all the bands.

\subsubsection{Fourier parameters}
\begin{figure*}
\begin{tabular}{cc}
\resizebox{1.0\linewidth}{!}{\includegraphics*{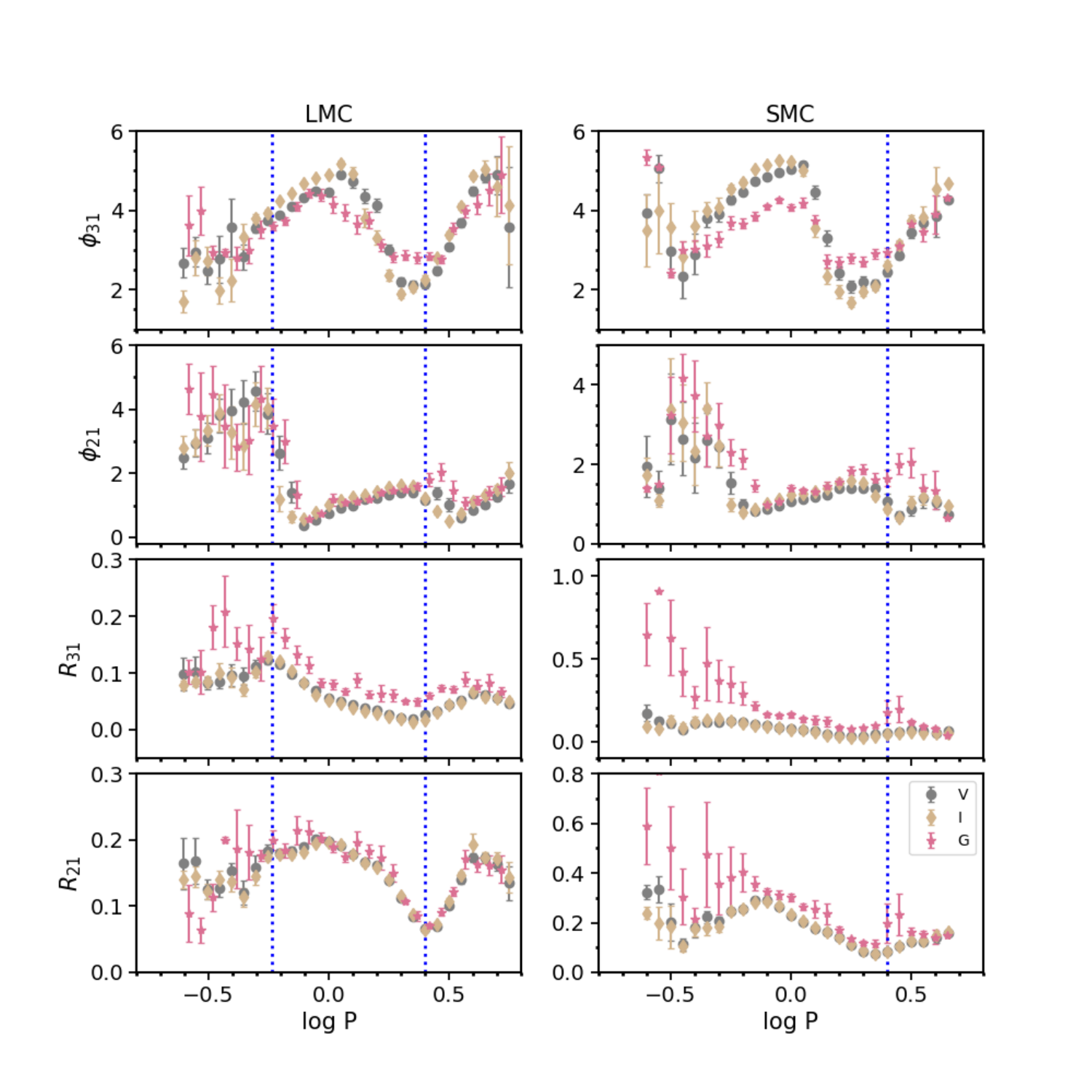}}& 
\end{tabular}
\caption{Fourier parameters for LMC (left panel) and SMC (right panel) FO Cepheids obtained from the observations in $V,I,G$-bands. The FPs shown are evaluated using sliding mean values with steps of $0.05$~dex in $\log{P}$ and a bin width of $0.1$~dex. The vertical dashed blue lines indicate $\log{P}=-0.23~(P=0.58~\rm d)$ and $\log{P}=0.4~(P=2.5~\rm d)$.}
\label{fig:fourier_lmcsmc_obs}
\end{figure*}

The Fourier parameters (FPs) of the light curves of FO Cepheids from the observations and the models for both 
LMC and SMC are obtained from the Fourier decomposition method. Fig.~\ref{fig:fourier_lmcsmc_obs} represents the FPs as a function of the period for FO Cepheids in the LMC and SMC. We have selected only $V, I, G$ bands for better visualisation because of the good phase coverage. Moreover, to distinguish the variation of the FPs as a function of period, we have sliced the distribution in terms of $\log{P}$ in steps of $0.05$~dex and a bin width of $0.09$~dex. With a series of experiments of many choices, we found that this combination produced the lowest amount of scatter between the consecutive points. The mean values of the FPs and $\log{P}$ in each slice are obtained. It is important to note that in Fig.~\ref{fig:fourier_lmcsmc_obs} the FPs are plotted as a function of period on a logarithmic scale. However, as mentioned in the previous section, we have maintained the discussion of the results of the FPs with the period in days. For $P>0.58$ d, the Fourier phase parameters of the FO Cepheids for the LMC and SMC obtained using the observations are found to be consistent. However, we observed a difference in the Fourier amplitude parameters for $P<0.58$ d. We observed that the Fourier amplitude parameters of the LMC FO Cepheids are comparatively smaller as compared to the SMC in this period range. We also emphasize that for near-infrared light curves, these differences in amplitude and phase parameters decreases.

The Fourier parameters for the LMC and SMC FO Cepheids in $V, I, G$ bands obtained from the observations 
are shown in Fig.~\ref{fig:fourier_lmcsmc_obs}. It can be seen from the figure that both $R_{21}$ and $R_{31}$ 
exhibit a change in the structure with the period, the variation is more pronounced near $P=0.58$~d and $P=2.5$~d in the case of LMC at all wavelengths. For both LMC and SMC, we observed a dip in $R_{21}$ and $R_{31}$ near 
$P=2.5$~d. This is consistent with the study of \citet{bhar16a}.  In both the MCs, the dip is larger in the case of $R_{21}$ as compared to $R_{31}$. For both LMC and SMC, we observed a discrepancy in $R_{21}$ and $R_{31}$ values between $G$-band and  $(V,I)$-bands for $P<0.58$~d. There is a negligible change in the Fourier amplitude parameters between $V$ and $I$- bands. Similarly, a change in the Fourier phase parameters can be seen at the same periods,  around $P=0.58$~d and $P=2.5$~d at all wavelengths for both LMC and SMC. small differences in $\phi_{21}$ and $\phi_{31}$ values are seen between $V$ and $I$ bands. For the $G$- band, a small variation in the phase parameters is seen for $P>1.0$~d. However, dissimilarity in the phase parameters among the bands is observed for $P<1.0$~d. 

\begin{figure*}
\begin{tabular}{cc}
\resizebox{1.0\linewidth}{!}{\includegraphics*{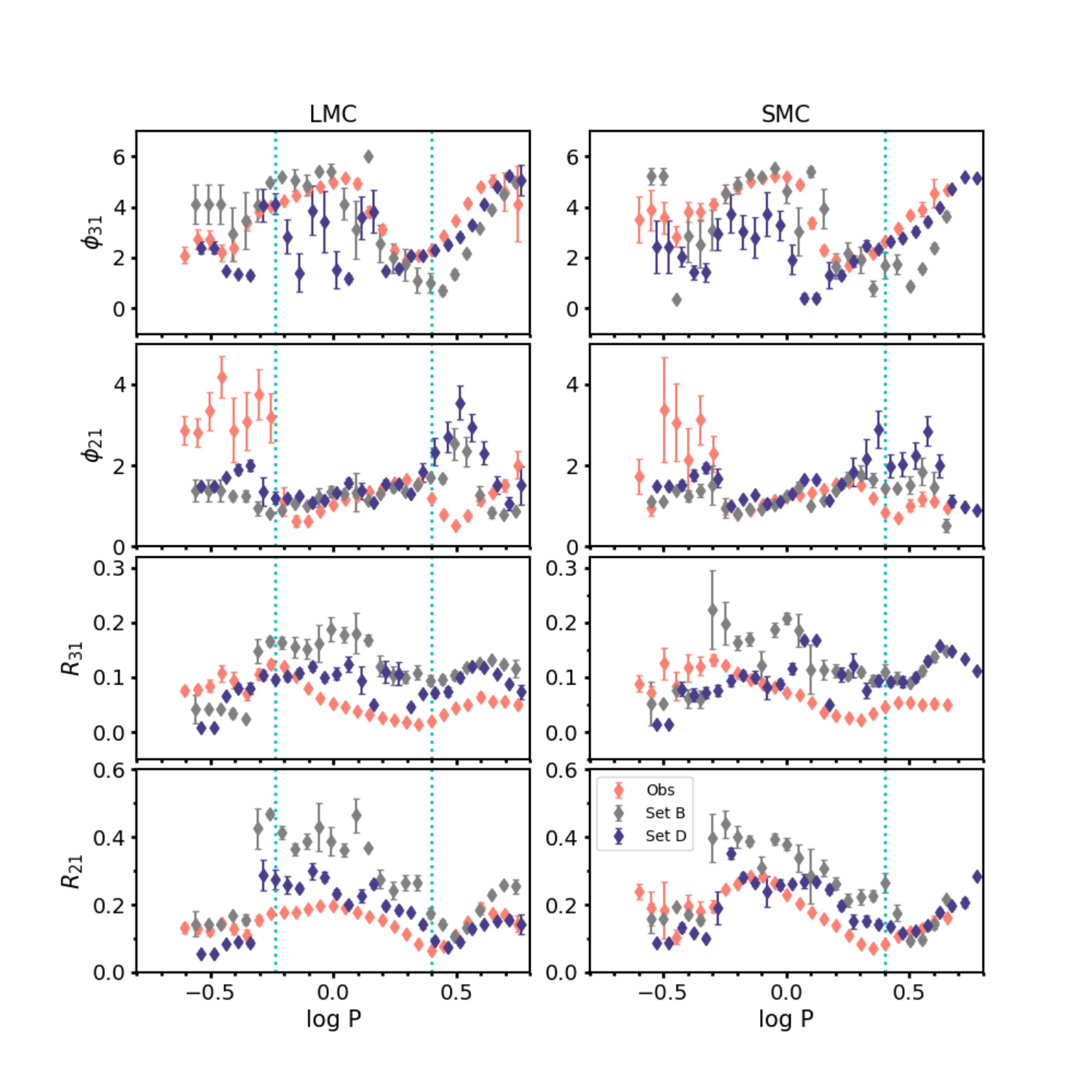}}& 
\end{tabular}
\caption{Fourier amplitude parameters for observed and theoretical light curves for FO Cepheids in the LMC and
SMC. Light orange colour represents the observations. Grey and blue colours represent sets B and D, respectively. The vertical  dashed lines (cyan) indicate $\log{P}=-0.23~(P=0.58~\rm d)$ and $\log{P}=0.4~(P=2.5~\rm d)$.}
\label{fig:fourier_lmcsmc_models}
\end{figure*}

The comparison of the FPs calculated using the observed data and theoretical models in the MCs is shown 
in Fig.~\ref{fig:fourier_lmcsmc_models} for the $I$-band. Light orange represents the FPs of the FO Cepheids 
obtained from the observations. Grey and blue colours represent the FPs of the FO models using sets B and D. From the figure, it can be seen that there is a change in the structure of amplitude parameters, $R_{21}$ and $R_{31}$, and the phase parameter, $\phi_{21}$ and $\phi_{31}$ at $P\sim2.5$~d. Comparatively, it is interesting to see that the amplitude parameters obtained using convection set D are more consistent with the observations than those obtained using set B in both bands. In terms of the phase parameters, $\phi_{21}$ obtained from the models have higher values when compared with the observed ones in the $I$-band but are found to be consistent with the observations in the $V$-band for both LMC and SMC. For both LMC/SMC, the $\phi_{31}$ parameter values are consistent with the observed ones. Observations clearly show a dip in $R_{k1}$ values around $P=2.5$~d, while the models using convection Set D models have such a dip near $P=2.8$~d. We have observed that there exists a large discrepancy between theoretical models and observations for the SMC.

For both the models and observations, we observed a change in the structure of the Fourier coefficients of FO Cepheids for both LMC and SMC. A thorough study of FO Cepheids by \citet{anto93,kien99, feuc00,buch04} from observational and theoretical perspectives explained that the $P_{1}/P_{4}=2$ resonance between the first and the fourth overtone is closely related with the changes in the structure of the Fourier coefficients. Moreover, the study by \citet{feuc00} clearly demonstrated that the inclusion of the convective energy transport is able to reproduce the observed behaviour of FO Cepheids in the Galaxy.

\subsubsection{Period-Colour relation}
Applying Stefan-Boltzmann law, the correlation between the Period-Colour (PC) and Amplitude-Colour (AC) 
relations at maximum and minimum light was explained by \citet{simo93}:
\begin{equation}
{\log T_{\max}-\log T_{\min}=\frac{1}{10}(V_{\max}-V_{\min})},
\label{equa:equa1}
\end{equation}
where $T_{\max}$ and $T_{\min}$ are the photospheric effective temperatures at maximum and minimum light, 
respectively. If $T_{\max}$ (hence the colour) is independent or weakly dependent on the period (i.e. the 
flat PC relation), Equation.~\ref{equa:equa1} predicts that there is a correlation between the amplitude in 
$V-$ band and the temperature (colour) at minimum light. The same also applies at maximum light.  

The colour terms ($V-\lambda$) at maximum, minimum, and mean light are defined as follows:
\begin{align}
(V-\lambda)_{\rm max} =& V_{\rm max} - \lambda_{\rm phmax}, \nonumber \\
(V-\lambda)_{\rm min} =& V_{\rm min} - \lambda_{\rm phmin}, \nonumber \\
(V-\lambda)_{\rm mean}=& V_{\rm mean} - \lambda_{\rm mean}.  
\label{eq:pcac}
\end{align}

Here $\lambda$ represents six photometric wavelengths in the optical, Gaia and near-infrared bands 
($IGG_{RP}JKs$). Also, $V_{\rm max}$, $V_{\rm min}$, $V_{\rm mean}$ denote the maximum, minimum and mean 
magnitudes in $V$-band. $\lambda_{\rm phmax}$ and $\lambda_{\rm phmin}$ represent the $\lambda-$band magnitude 
corresponding to the phase of maximum and minimum light in $V$-band. $\lambda_{\rm mean}$ is the mean magnitude in $IGG_{RP}JKs$- bands. 

\begin{figure*}
\begin{tabular}{cc}
\resizebox{1.0\linewidth}{!}{\includegraphics*{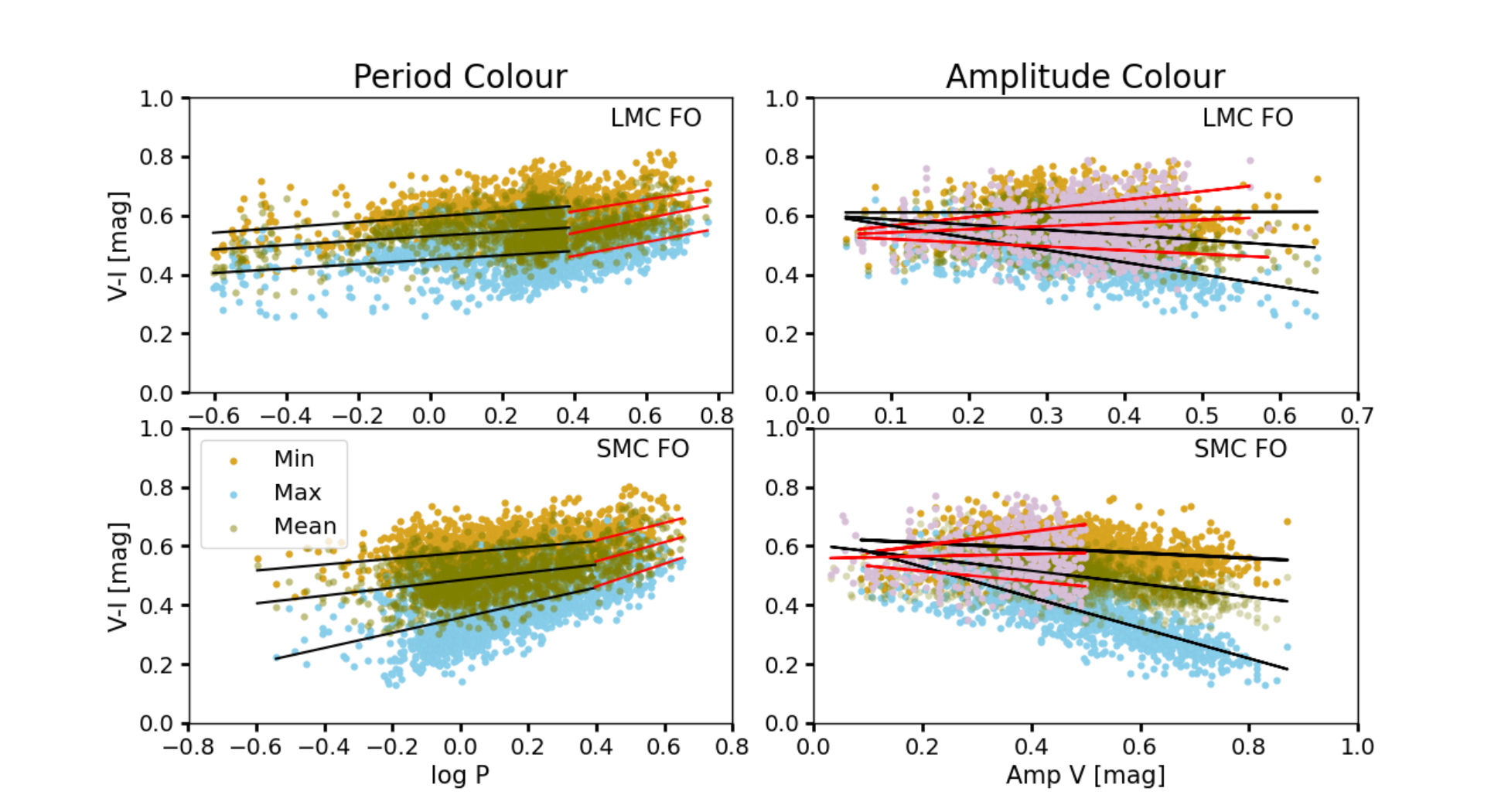}}& 
\end{tabular}
\caption{\textit{Left Panel}: PC relations for LMC and SMC FO Cepheids using $V,I$-bands. Sky blue, light purple and
orange represent the PC relations at maximum/minimum/mean light. \textit{Right Panel}: Same as the left panel but 
for AC. Navy blue represents the star with P > 2.5 d in both cases. In all the plots, solid black/red lines 
represent the linear regression to the PC/AC relations for FO Cepheids below and above 2.5 d, respectively. }
\label{fig:pcac_lmcsmc_fufo_ogle}
\end{figure*}

In this section, we discuss the results of the PC relations using six colour indices. The analysis is carried 
out for both FO and short-period FU Cepheids in the LMC and SMC at maximum, minimum, and mean light. 

The results of the $F$- test of the PC relations at mean light for FO and FU Cepheids at the obtained 
break-points determined using piecewise regression for both LMC and SMC are summarised in Tables \ref{tab:fvalue_pc_lmcsmc} and \ref{tab:fvalue_pc_fu_lmcsmc} in Appendix \ref{sec:ftest_pw}. The first ($P= 2.5$ d) and second ($P= 0.58$ d) rows in these tables indicate the results of the $F$- test obtained for the break-points as reported in the literature assuming that these break-points exist across all colour indices. Inclusion of the $F$- test results corresponding to these two break-points is for comparative purposes only. The remaining rows represent the $F$- test result for break-points obtained using piecewise regression. 
The locations of the break-points identified using piecewise regression are found to be different across different colour indices. Some of the break-points obtained are found to occur near $P=1.0$~d and $P=2.5$~d in the case of LMC and SMC FO Cepheids. These break-points are found to be significant and some of them are found to show higher confidence levels as compared to the break-point exactly at $P=2.5$~d as reported in the literature. Similarly, the break-points near $P=2.5$~d has been observed across all colour indices for both LMC and SMC FU Cepheids. Athough, the break-points are found to be not significant in the case of LMC FU Cepheids, the $F$-test results show that the break-points are significant in the case of SMC FU Cepheids across different colour indices. Break-points around $P=4.0$~d have been observed across all colour indices for the LMC FU Cepheids, although, these break-points are found to be not significant. It can be seen from Tables \ref{tab:fvalue_pc_lmcsmc} and \ref{tab:fvalue_pc_fu_lmcsmc} that certain break-points obtained by piecewise regression are significant, whereas some are not.

Table \ref{tab:pc_fufo_mean_lmcsmc} summarised the coefficients of the PC relations and the results of the statistical $F$- test for the LMC and SMC FU/FO Cepheids using the observations at mean light for short periods ($P<2.5$~d), long periods ($P>2.5$~d) and all periods. Investigation of the significance of the break-points
in the PC relations for LMC FO Cepheids near $P=2.5$~d has been carried out for stars with $P>0.58$ d. It is evident from Table \ref{tab:pc_fufo_mean_lmcsmc} that for FO Cepheids, the slopes of the PC relations in the SMC are steeper than the LMC counterparts at mean light. This is true for all/short/long periods. Similarly, the same holds true for maximum and minimum light. On the other hand, the slopes of the PC relations for LMC FU Cepheids at mean light are found to be steeper than the SMC for all periods. However, the slopes of the PC relations for short and long periods in the SMC are seen to be steeper than the LMC counterparts. This is true for maximum and minimum light.

Fig.~\ref{fig:ftest_pc} in the Appendix~\ref{sec:f_test_pcpl} depicts the probability of the $F$- test of the PC relations at maximum/minimum/mean light for FU Cepheids (upper panel) and FO Cepheids (lower panel) in both LMC and SMC. The upper panel shows that for the LMC FU Cepheids, the $F$-test is not significant for $V-G$ and $V-Y$ colours at maximum and minimum light, respectively. In the case of SMC, the $F$- test is found to be not significant for $V-I$ at minimum light and $V-G$ at maximum light. On the other hand, the lower panel of Fig.~\ref{fig:ftest_pc} in the Appendix~\ref{sec:f_test_pcpl} shows that for the FO Cepheids in the LMC, the $F$-test results are neither significant at maximum nor minimum light for $V-G$ and $V-J$. Similarly, the $F$- test is found to be not significant for $V-G_{RP}$ at minimum light and $V-Y$ at maximum light. For the SMC, the $F$- test results are neither significant at maximum nor minimum light for $V-G_{RP}$ and $V-Y$. Moreover, the $F$-test results are not significant for $V-J$ and $V-Ks$ at maximum light.  

The observed behaviour of the PC relations at mean light is the average over all pulsation phases. To explain the significance of the $F$-test at mean light when the $F$- test is not significant at either maximum or minimum light, we obtain the PC relation for $100$ pulsation phases with  $V-Y$ as the colour index 
for SMC FO Cepheids following the steps described in \citet{kurb23}. The probability of the $F$-test is shown in Fig.~\ref{fig:probability_ftest} in Appendix~\ref{sec:f_test_pcpl}. From the calculation we found that there exist more pulsational phases ($52$ phases) where the results of the $F$-test are significant which contribute to the significance in the PC relation at mean light.

The left panel of Fig.~\ref{fig:pcac_lmcsmc_fufo_ogle} displays the empirical PC relations of LMC and SMC FO Cepheids using $(V, I)$ photometric bands with a break-point near $P=2.5$~d , respectively. In 
the case of SMC, the relations is flatter only at minimum light, and we find a slope PC relations at maximum and mean light. For stars with periods greater than 2.5 d, we get a slope PC relations at maximum/mean/minimum light for both LMC and SMC. It is important to note that the PC relations of the LMC FO Cepheids at maximum and minimum light are nearly parallel, which is not observed in the case of SMC FO Cepheids. This is consistent with the results found by \citet{bhar16} using OGLE-III database. 

From the theoretical perspective, in the case of LMC, the models with $P > 0.58$ d were selected to maintain consistency with observational data. Using the models, the presence of a break-point in the PC relations of FO Cepheids at $P=2.5$ d was examined for both the LMC and SMC. An identical methodology was employed for the analysis of the AC and PL relations. Using sets B and D, we found that $V-I, V-J$ and $V-Ks$ exhibit a break-point in the PC relations for the LMC FO Cepheids at mean light.  However, the break-point was found to be not significant for $V-G$ and $V-G_{RP}$ in both cases. In addition, the break-point is significant for all colour indices using Set B at maximum light in case of the LMC. At minimum light, the break-point is significant for $V-I$ and $V-G$. On the other hand, the break-point is found to be significant only for $V-J$ and $V-Ks$ at minimum light. For the SMC, a break-point was observed in all color indices at maximum light, whereas at minimum light, the break-points are significant only for $V-G$, $V-J$, and $V-Ks$. 

The coefficients of the PC relations and results of the $F$- test using sets B and D for LMC and SMC are available in the supplementary data. The results from the $t$- test suggest that the PC relation obtained using set B is more consistent with the observations for both LMC and SMC. The results of the $t$- test are summarised in the supplementary data. Similarly, we also obtained a break-point in the PC relations for FU Cepheids using convection set B for both LMC and SMC. The coefficient of the PC relations for FU Cepheids is available in the supplementary data. The results from the $t$- test suggest that the PC relations obtained using set B is more consistent with the observations for both LMC and SMC. The results of the $t$- test are summarised in the supplementary data. Similarly, we also obtained a break-point in the PC relations for FU Cepheids using convection set B for both LMC and SMC. The coefficient of the PC relations for FU Cepheids is available in the supplementary data.

\begin{table*}
\begin{center}
\caption{Coefficients of the PC relations using the LMC and SMC FU/FO Cepheids and $F$-test results to test the significance of the break-points detected near $P=2.5$~d from piecewise regression using the observed data at mean light. The subscripts $_{\rm s}$, $_{\rm l}$ and $_{\rm all}$ represents short ($\sim P<2.5$~d), long ($\sim P>2.5$~d) and all periods, respectively. The bold-faced entries indicate the significance of the break-points in the PC relations. The numbers given within the parentheses are the break-points in days obtained using piecewise regression analysis.}
\begin{tabular}{c c c c c c c c c c c c } \\ \hline \hline
Type&    &  $a_{\rm all}$ & $b_{\rm all}$ & $a_{\rm s}$ & $b_{\rm s}$ & $a_{\rm l}$ & $b_{\rm l}$ &  $F$ & $p(F)$    \\ \hline 
\multicolumn{9}{c}{LMC}\\ \hline
FU 
&$V-I$~($2.523$)  & $0.219\pm0.007$  & $0.535\pm0.004$   & $0.150\pm0.032$
& $0.55\pm0.010$  & $0.215\pm0.008$    & $0.538\pm0.005$   &  
$2.790$    & $0.061$ \\
&$V-G$~($2.483$)  & $0.177\pm0.010$   & $-0.013\pm 0.006$  & $0.211\pm 0.060$  & $-0.018\pm0.018$   & $0.185\pm0.012$ 
&  $-0.018\pm0.007$  & $0.915$   & $0.400$ \\
&$V-G_{RP}$~($2.483$) & $0.242\pm0.011$  & $0.456\pm0.006$   & 
$0.172\pm0.062$  & $0.469\pm0.019$  & $0.233\pm0.013$  &
$0.462\pm0.008$   & $1.840$   & $0.159$  \\ 
&$V-Y$~($2.535$)  & $0.285\pm0.013$  & $0.752\pm0.007$   & $0.218\pm0.064$  &
$0.769\pm0.020$   & $0.283\pm0.016$  & $0.754\pm0.010$   & 
$0.697$   & $0.498$ \\
&$V-J$~($2.494$)  & $0.352\pm0.014$  & $0.922\pm0.008$   & $0.319\pm0.074$  & $0.931\pm0.022$
& $0.352\pm0.018$  &  $0.923\pm0.011$  & $0.103$   & $0.902$ \\
&$V-Ks$~($2.423$) & $0.484\pm0.016$  & $1.198\pm0.009$   & $0.359\pm0.080$  &
$1.225\pm0.025$     & $0.478\pm0.019$  &  $1.203\pm0.012$  & 
$1.743$   & $0.175$ \\  \hline
FO 
&$V-I$~($2.440$)  & $0.061\pm0.008$   & $0.534\pm0.002$   & $0.028\pm0.012$   & $0.541\pm0.003$ & $0.243\pm0.028$ & $0.442\pm0.014$  & $19.324$ & \textbf{0.000} \\ 
&$V-G$~($2.466$)  & $0.022\pm0.005$   & $0.066\pm0.002$   & $0.013\pm0.007$   & $0.069\pm 0.002$  & $0.088\pm0.019$ 
& $0.033\pm0.009$   & $6.681$   & \textbf{0.000} \\
&$V-G_{RP}$~($2.540$) & $0.074\pm0.010$   & $0.474\pm0.003$   & $0.043\pm0.015$   & $0.479\pm0.004$   & $0.229\pm0.040$  & $0.396\pm0.021$   & $8.576$   & \textbf{0.000} \\
&$V-Y$~($2.434$)  & $0.099\pm0.012$  & $0.734\pm0.004$   & $0.042\pm0.019$  & $0.744\pm0.005$ & $0.351\pm0.048$  
& $0.609\pm0.025$   & $16.248$   & \textbf{0.000} \\
&$V-J$~($2.449$)  & $0.123\pm0.014$  & $0.901\pm0.005$   & $0.065\pm0.022$  & $0.912\pm0.005$ & $0.438\pm0.055$  & $0.742\pm0.029$  & $16.572$   & \textbf{0.000} \\
&$V-Ks$~($2.454$) & $0.183\pm0.017$  & $1.181\pm0.006$   & $0.115\pm0.026$  & $1.194\pm0.006$ & $0.598\pm0.026$  & $0.970\pm0.006$   & $19.533$   & \textbf{0.000} \\ \hline 
\multicolumn{9}{c}{SMC}\\ \hline
FU
& $V-I$~($2.654$) &  $0.182\pm0.004$ & $0.564\pm0.002$   & $0.190\pm0.013$ & $0.563\pm0.003$  & $0.226\pm0.014$  & $0.533\pm0.009$   & 
$6.146$  & \textbf{0.000} \\
& $V-G$~($2.517$) & $0.119\pm0.005$  & $0.043\pm0.003$   & $0.131\pm0.009$  & 
$0.039\pm0.003$  & $0.061\pm0.016$  & $0.086\pm0.012$
& $5.754$   & \textbf{0.000} \\
& $V-G_{RP}$~($2.409$)    & $0.224\pm0.007$  & $0.468\pm0.003$   & $0.238\pm0.013$  &
$0.464\pm0.004$   & $0.158\pm0.035$  & $0.517\pm0.026$   & 
$16.611$  & \textbf{0.000} \\
& $V-Y$~($2.322$) & $0.279\pm0.011$  & $0.713\pm0.004$   & $0.278\pm0.038$  & 
$0.716\pm0.008$   & $0.338\pm0.025$  & $0.675\pm0.015$  & $3.309$   & \textbf{0.000} \\
& $V-J$~($2.371$) & $0.296\pm0.011$  & $0.985\pm0.004$   & $0.339\pm0.035$  &
$0.980\pm0.007$   & $0.374\pm0.025$  & $0.932\pm0.015$  &
$7.300$   & \textbf{0.000} \\
& $V-Ks$~($2.275$)    & $0.420\pm0.012$  & $1.241\pm0.005$   & $0.479\pm0.043$  & 
$1.234\pm0.009$   & $0.500\pm0.028$  & $1.187\pm0.028$  &
$6.957$   & \textbf{0.000} \\ \hline
 FO 
&$V-I$~($2.494$)  & $0.160\pm0.007$  & $0.485\pm0.002$   & $0.134\pm0.009$  & $0.485\pm0.002$
& $0.279\pm0.083$  & $0.441\pm0.041$   & $9.656$ & \textbf{0.000} \\
& $V-G$~($2.506$) & $0.040\pm0.004$   & $0.064\pm0.001$   & $0.026\pm0.005$   & $0.064\pm0.001$   & $0.099\pm0.041$ 
& $0.041\pm0.020$   & $8.597$   & \textbf{0.000} \\ 
& $V-G_{RP}$~($2.454$)    & $0.164\pm0.010$  & $0.421\pm0.002$   & $0.136\pm0.013$  & $0.423\pm0.003$ 
& $0.292\pm0.084$  & $0.371\pm0.041$   & $5.856$   & \textbf{0.008} \\
&$V-Y$~($2.296$)  & $0.223\pm0.014$  & $0.625\pm0.003$   & $0.177\pm0.020$  & $0.626\pm0.036$
& $0.347\pm0.102$  & $0.583\pm0.048$   & $6.058$   & \textbf{0.009} \\
&$V-J$~($2.338$)  & $0.284\pm0.016$  & $0.843\pm0.003$   & $0.226\pm0.022$  & $0.844\pm 0.004$    & $0.508\pm0.118$  & $0.757\pm0.056$   & $8.116$   & \textbf{0.000} \\
& $V-Ks$~($2.344$)    & $0.409\pm0.019$  & $1.068\pm0.004$   & $0.344\pm0.026$  & $1.070\pm0.004$
& $0.672\pm0.148$  & $0.967\pm0.070$   & $6.806$   & \textbf{0.000} \\   \hline \hline
\end{tabular}
\label{tab:pc_fufo_mean_lmcsmc}
\end{center}
\end{table*}

\subsubsection{Amplitude-Colour relation}

This subsection discusses the amplitude-colour (AC) relations of FO and FU Cepheids in the LMC and 
SMC. Table \ref{tab:ac_fufo_mean_lmcsmc} summarises the coefficients of the AC relations for the FU/FO Cepheids in the LMC and SMC at mean light for all/short/long-periods obtained from the observed data. The results of the statistical $F$-test are also summarised in Table ~\ref{tab:ac_fufo_mean_lmcsmc}. It can be seen from Table ~\ref{tab:ac_fufo_mean_lmcsmc} that the break-point in the AC relations at $P=2.5$~d is significant in all the colours for both FU/FO Cepheids in the MCs. The same is true at maximum and minimum light. 

Using the OGLE-IV ($V, I$) photometric bands, the AC relations of FO Cepheids in the LMC and SMC are 
displayed in the right-hand panel of Fig.~\ref{fig:pcac_lmcsmc_fufo_ogle}. At maximum light, the AC relations 
has a negative slope for both LMC and SMC for periods less than $2.5$~d. However, at minimum light, we observe a flat AC relations for both LMC and SMC. For FO Cepheids with periods greater than $2.5$~d, the AC relations displays a negative slope at maximum light but positive at minimum light for both LMC and SMC. The negative slope at maximum light indicates that higher-amplitude stars are bluer whereas the positive slope at minimum light indicates that higher-amplitude stars are redder \citep{bhar14,bhar16a}. It is evident from Table \ref{tab:ac_fufo_mean_lmcsmc} that there exists a break-point in the AC relations of the FO and FU Cepheids at for both LMC and SMC at $P=2.5$~d.

Using theoretical light curves, the AC relations obtained using convection sets B and D exhibits a break-point at 
$P=2.5$~d for both FO/FU Cepheids in the LMC and SMC. The coefficients of the AC relations and results of the statistical $F$ -test obtained using sets B and D for the LMC and SMC are available in the supplementary data. We have also conducted a $t$-test to compare the slopes of the AC relations obtained using sets B and D with the observations. In contrast to the PC relations, we found that the AC relation obtained using set D are more consistent with the observations for both LMC and SMC.

\begin{table*}
\begin{center}
\caption{Coefficients of the AC relations using LMC and SMC FU/FO Cepheids and the $F$- test results to test the significance of the break-point at $P=2.5$ d at mean light using the observations. The subscripts s, l and all represents short ($P< 2.5$ d), long ($P>2.5$ d) and all periods, respectively. The bold-faced entries indicate the break-point in the AC relations }
\begin{tabular}{c c c c c c c c c c c c c} \\ \hline \hline
Type&   &  $a_{\rm all}$ & $b_{\rm all}$ & $a_{\rm s}$ & $b_{\rm s}$ & $a_{\rm l}$ & $b_{\rm l}$ &  $F$ & $p(F)$    \\ \hline 
\multicolumn{9}{c}{LMC}\\ \hline
FU 
& $V-I$         & $-0.150\pm0.005$ & $0.773\pm0.004$   & $-0.098\pm0.016$  & $0.676\pm0.013$ 
& $-0.147\pm0.005$   & $0.776\pm0.004$   & $118.703$ & \textbf{0.000} \\ 
& $V-G$         & $-0.039\pm0.003$  & $0.138\pm0.002$   & $-0.018\pm0.011$  & $0.103\pm0.009$  
& $-0.039\pm0.003$ & $0.140\pm 0.002$  & $32.412$  & \textbf{0.000}  \\ 
& $V-G_{RP}$    & $-0.118\pm0.007$ & $0.691\pm0.006$   & $-0.099\pm0.023$  & $0.622\pm0.019$ 
& $-0.113\pm0.007$ & $0.693\pm0.005$   & $77.287$  & \textbf{0.000} \\
& $V-Y$         & $0.739\pm0.012$  & $-0.170\pm0.034$  & $0.750\pm0.014$  & $-0.230\pm0.039$ 
& $0.695\pm0.022$  & $0.049\pm0.064$   & $22.127$  & \textbf{0.000} \\
& $V-J$         & $0.804\pm0.010$  & $-0.130\pm0.028$  & $0.821\pm0.011$  & $-0.205\pm0.033$ 
& $0.745\pm0.018$  & $0.129\pm0.052$   & $33.159$  & \textbf{0.000} \\ 
& $V-Ks$        & $-0.251\pm0.041$ & $1.332\pm0.014$   & $-0.372\pm0.047$ & $1.358\pm0.017$
& $0.180\pm0.069$  & $1.233\pm0.024$   & $44.332$  & \textbf{0.000} \\ 
\hline
FO
& $V-I$      & $-0.109\pm0.017$    & $0.589\pm0.006$   & $-0.167\pm0.019$ & $0.603\pm0.007$ 
& $0.094\pm0.030$   & $0.538\pm0.010$   & $72.274$  & \textbf{0.000} \\ 
& $V-G$      & $-0.025\pm0.010$ & $0.082\pm0.004$   & $-0.048\pm0.012$  & $0.088\pm0.004$   & $0.041\pm0.019$
& $ 0.063\pm0.007$  & $12.810$  & \textbf{0.000} \\
& $V-G_{RP}$ & $-0.100\pm0.022$ & $0.530\pm0.008$   & $-0.174\pm0.026$ & $0.549\pm0.009$
& $0.123\pm0.038$  & $0.472\pm0.013$   & $34.439$  & \textbf{0.000} \\
& $V-Y$      & $-0.130\pm0.028$ & $0.804\pm0.010$   & $-0.205\pm0.033$ & $0.821\pm0.011$ 
& $0.129\pm0.052$  & $0.745\pm0.018$   & $33.159$  & \textbf{0.000} \\ 
& $V-J$      & $-0.138\pm0.029$ & $0.773\pm0.010$   & $-0.208\pm0.034$ & $0.787\pm0.012$ 
& $0.107\pm0.053$  & $0.719\pm0.018$   & $31.565$  & \textbf{0.000} \\
& $V-Ks$     & $-0.255\pm0.041$ & $1.329\pm0.014$   & $-0.393\pm0.047$ & $1.361\pm0.017$ 
& $0.222\pm0.069$  & $1.215\pm0.024$   & $50.362$  & \textbf{0.000} \\ \hline
	\multicolumn{9}{c}{SMC}\\ \hline
FU 
& $V-I$      & $-0.179\pm0.007$ & $0.715\pm0.004$   & $-0.144\pm0.008$ &  $0.674\pm0.005$
& $-0.188\pm0.011$ & $0.757\pm0.006$   & $136.503$ & \textbf{0.000}\\
& $V-G$      & $-0.108\pm0.005$ & $0.691\pm0.005$   & $-0.083\pm0.007$  & $0.646\pm 0.007$
& $-0.104\pm0.006$ & $0.717\pm0.006$   & $126.242$ & \textbf{0.000} \\
& $V-G_{RP}$ & $-0.108\pm0.006$ & $0.663\pm0.005$   & $-0.082\pm0.008$  & $0.617\pm0.008$ 
& $-0.104\pm0.007$ & $0.690\pm0.006$   & $130.059$ & \textbf{0.000} \\
& $V-Y$      & $-0.152\pm0.010$ & $0.392\pm0.009$   & $-0.114\pm0.012$ & $0.327\pm0.012$
& $-0.169\pm0.015$ & $0.462\pm0.013$   & $102.203$ & \textbf{0.000} \\
& $V-J$      & $-0.173\pm0.010$ & $0.710\pm0.010$   & $-0.128\pm0.013$ & $0.635\pm0.012$
& $-0.193\pm0.015$ & $0.793\pm0.013$   & $127.896$ & \textbf{0.000} \\
& $V-Ks$     & $0.219\pm0.010$  & $1.087\pm0.010$   & $-0.153\pm0.013$ & $0.982\pm0.012$ 
& $-0.259\pm0.014$ & $1.207\pm0.013$   & $243.133$ & \textbf{0.000} \\ \hline
FO 
& $V-I$         & $-0.236\pm0.010$ & $0.618\pm0.005$   & $-0.229\pm0.010$ & $0.611\pm0.005$ 
& $0.034\pm0.037$   & $0.563\pm0.013$   & $54.260$  & \textbf{0.000} \\ 
& $V-G$         & $-0.046\pm0.005$  & $0.090\pm0.002$   & $-0.037\pm0.006$  & $0.084\pm0.003$  
& $0.007\pm0.023$   & $0.088\pm0.008$   & $27.491$  & \textbf{0.000} \\ 
& $V-G_{RP}$    & $-0.175\pm0.014$ & $0.526\pm0.007$   & $-0.156\pm0.015$ & $0.512\pm0.007$ 
& $0.079\pm0.052$   & $0.488\pm0.019$   & $43.742$  & \textbf{0.000} \\ 
& $V-Y$         & $-0.302\pm0.020$ & $0.794\pm0.009$   & $-0.279\pm0.021$ & $0.776\pm 0.010$ 
& $0.037\pm0.076$   & $0.743\pm0.026$   & $36.749$   & \textbf{0.000} \\ 
& $V-J$         & $-0.370\pm0.023$ & $1.052\pm0.010$   & $-0.340\pm0.024$ & $1.028\pm0.011$ 
& $0.077\pm0.087$   & $0.987\pm0.030$   & $49.930$  & \textbf{0.000} \\
& $V-Ks$        & $-0.508\pm0.028$ & $1.358\pm0.013$   & $-0.472\pm0.029$ & $1.328\pm0.014$
& $0.147\pm0.104$  & $1.250\pm0.036$   & $62.021$  & \textbf{0.000} \\ \hline  \hline
\end{tabular}
\label{tab:ac_fufo_mean_lmcsmc}
\end{center}
\end{table*}

\subsubsection{Period-Luminosity Relation}

Using piecewise regression, we obtained multiple break-points in the PL relations of LMC and SMC FO/FU Cepheids at mean light. Similar to the PC relations, the locations of the break-points in the PL plane are found to be different across different photometric bands. The significance of break-points obtained from the piecewise regression analysis is studied using the statistical $F$-test and the results are provided in Tables~\ref{tab:fvalue_pl_lmcsmc} and \ref{tab:fvalue_pl_fu_lmcsmc} in Appendix~\ref{sec:ftest_pw}. Similar to the PC relations, the first and second rows ($P = 2.5$ d and $P = 0.58$ d) in these tables indicate the results of the $F$- test obtained for the break-points as reported in the literature assuming that the break-points exist across all photometric bands. Inclusion of the $F$- test results corresponding to these two break-points from the literature is for comparative purposes only.

It is evident from Tables~\ref{tab:fvalue_pl_lmcsmc} and \ref{tab:fvalue_pl_fu_lmcsmc} that the locations of some of the break-points identified using piecewise regression analysis are found near $P=2.5$ d for both FO and FU Cepheids in the LMC/SMC. These break-point locations emerge prominently across all photometric bands for both LMC and SMC and some of them are found to have higher confidence levels as compared to the break-point exactly at $P=2.5$~d as reported in the literature. These break-points near $P=2.5$~d across all bands are found to be significant in the case of LMC FO Cepheids and SMC FO/FU Cepheids. The break-points detected for the LMC FU Cepheids around $P=2.5$~d for all photometric bands are found to be significant. Break-points near $P=4.0$~d using the LMC FU Cepheids has also been observed across all the bands and some of the break-points are found to be significant ($G$ and $Y$ bands). We also observed a break-point around $P=1.0$~d in the case of SMC FO Cepheids. However, these break-points are found to be not significant when $F$- test is carried out. It is evident from Tables \ref{tab:fvalue_pl_lmcsmc} and \ref{tab:fvalue_pl_fu_lmcsmc} that certain break-points found from the piecewise regression are not significant when $F$- test is applied.

\begin{figure}
\begin{tabular}{cc}
\resizebox{1.0\linewidth}{!}{\includegraphics*{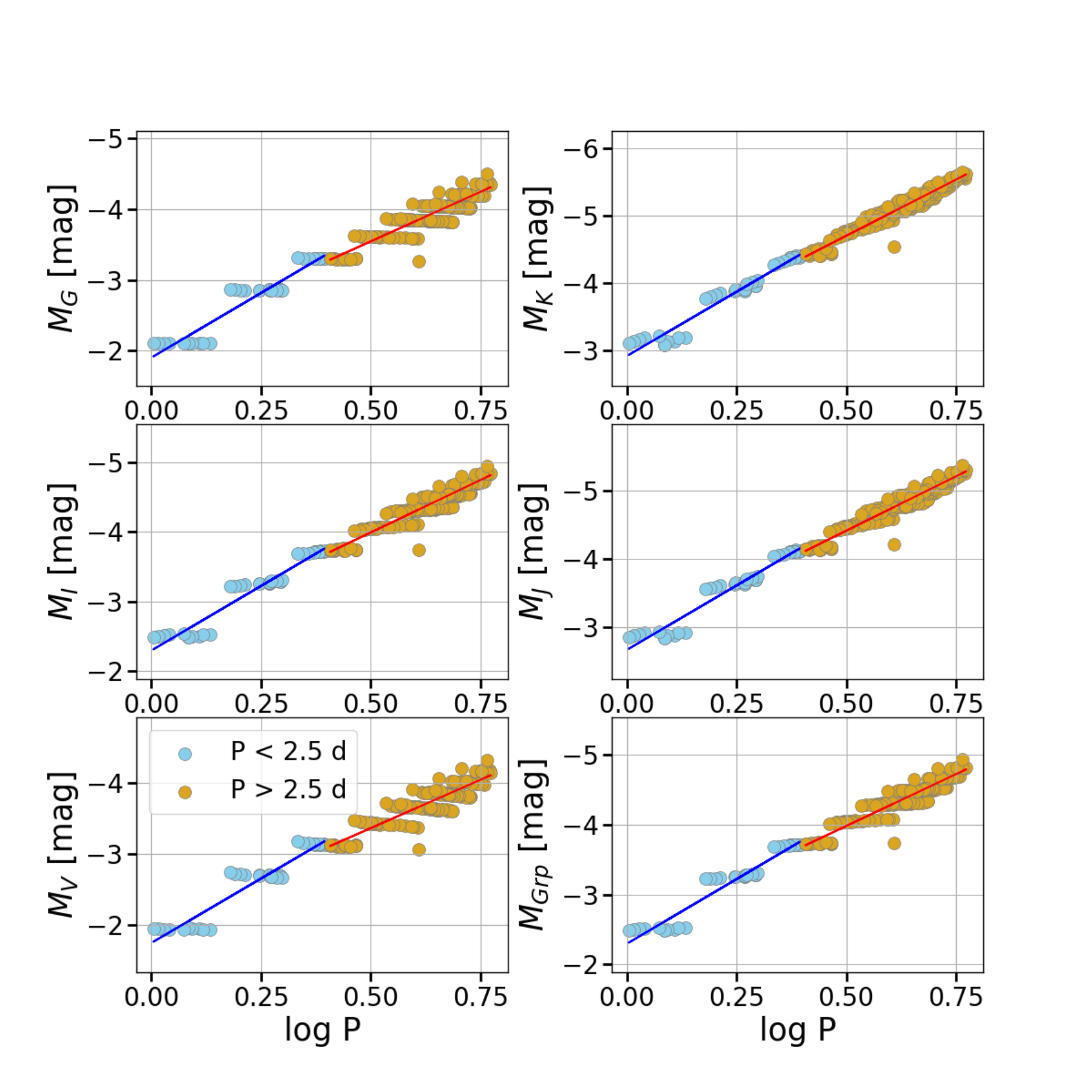}}& 
\end{tabular}
\caption{PL relations for LMC FO Cepheids using Set D with period $P>0.58$ d in $VIGG_{RP}JKs$ photometric bands. Blue represents the models with $P<2.5$ d and orange represents the models with $P>2.5$ d. The significance of the break-point at $P=2.5$ d is determined using the statistical $F$-test.}
\label{fig:pl_plot_lmc_setd}
\end{figure}

The coefficients of the PL relations and the result of the $F$- test at approximately $P=2.5$~d for FO/FU Cepheids in the LMC and SMC at mean light obtained using the observations are summarised in Table \ref{tab:slopes_pl_maxmin_lmcsmc} for all/short/long periods. Fig.~\ref{fig:pl_plot_lmc_setd} represents the PL relations at mean light obtained using convection set D at multiwavelengths. 
Fig.~\ref{fig:ftest_pl} in Appendix~\ref{sec:f_test_pcpl} depicts the probability ($p(F)$) 
of the $F$- test of the PL relations for various photometric bands at maximum, minimum, and mean light. It is evident from the upper panel of Fig.~\ref{fig:ftest_pl} and Table~\ref{tab:slopes_pl_maxmin_lmcsmc} that for the LMC FU Cepheids, the break-point in the mean light PL relations near $P=2.5$~d is not significant. However, we observed a break-point in the PL relations near $P=2.5$~d for FU Cepheids in the SMC for all bands. In addition, we confirm a break-point in the PL relations of the LMC and SMC FO Cepheids near $P=2.5$~d. For the SMC FO Cepheids, the break-point in the PL relations is not significant for $G$ and $Y$ at minimum light.

However, the break-point is significant at both maximum and minimum light in the case of SMC FU Cepheids.
For LMC, the break-point at $P=2.5$ d investigated at maximum light is found to be significant in the $J$ and $Ks$ bands using set B, and across all bands using set D. However, at minimum light, the break-point is significant in all bands for both sets B and D. Similarly, for the SMC, the break-point is significant at both maximum and minimum light using set B. For Set D, the break-point is significant at minimum light in all the bands, however the break-point is significant for $VIJKs$ at maximum light. Furthermore, for both LMC and SMC, the PL relations exhibits a break-point at $P=2.5$ d at mean light for both sets B and D.  The coefficients of the PL relations obtained using convection sets B and D are given in the supplementary data.

The results of the coefficients and the $t$-test of the PL relations of the FO Cepheids in the LMC and SMC are
summarised in Tables~\ref{tab:pl_models_lmc} and \ref{tab:pl_models_smc} in Appendix~\ref{sec:t_test_pl_appen}. 
For the LMC models, the PL slopes using set B (without turbulent pressure and turbulent flux) exhibit
slopes similar to the empirical ones compared to convection set D. In $IJKs$- bands, the PL slopes of the models using sets B and D seem more consistent with the empirical ones. We observed that the slopes of the 
PL relations obtained using the observations and convection sets B and D are the steepest in the $Ks$ band. The results of the $t$-test are also summarised in the supplementary data. 

Similarly, for the SMC, the PL slopes obtained using convection set B display similar PL slopes as the 
empirical ones compared to the PL slopes obtained using set D. In $JKs$-bands, it can be seen that the slope of 
the PL relations obtained using set D is more consistent with the observations. In the optical bands, the 
PL slopes obtained using sets B and D are similar to the empirical ones with a difference  $<10\%$. 

It is worth noting that the dispersion of the theoretical PL slopes decreases significantly as we go from the 
optical to the near-infrared wavelengths for both LMC and SMC. Furthermore, the slopes of the PL relations become steeper with increasing in wavelength. This is similar to the results in the PL relations of BL Her stars and RR Lyraes obtained in \citet{neel17, beat18, bhar20, das21, das24}.

\begin{table*}
\begin{center}
\caption{Coefficients of the PL relations using the LMC and SMC FU/FO Cepheids and the $F$- test results to test significance of the break-points detected near $P=2.5$ d using piecewise regression at mean light from the observed data. The subscripts s, l and all represents short ($\sim P< 2.5$ d), long ($\sim P>2.5$ d) and all periods, respectively. The bold-faced entries represent the significance of the break-point in the PL relations. The numbers given within the parentheses are the break-points in days obtained using piecewise regression analysis.}
\begin{tabular}{c c c c c c c c c c c c c c} \\ \hline \hline
Type  & Band & $a_{\rm all}$ & $b_{\rm all}$ & $a_{\rm s}$ & $b_{\rm s}$ & $a_{\rm l}$ & $b_{\rm l}$ & $F$ & $p(F)$ \\ \hline 
\multicolumn{10}{c}{LMC} \\ \hline
FU 
& $V$~($2.540$)       & $-2.762\pm0.021$  & $17.210\pm0.013$  &
$-2.905\pm0.108$  & $17.239\pm0.034$  & $-2.780\pm0.026$ 
& $17.222\pm0.016$  & $2.145$   & $0.117$\\
& $I$~($2.612$)       & $-2.982\pm0.015$  & $16.677\pm0.009$  &
$-3.066\pm0.080$  & $16.690\pm0.026$  & $-3.004\pm0.019$
& $16.693\pm0.012$  & $3.275$   & $\textbf{0.038}$\\ 	
& $Y$~($2.482$)       & $-3.030\pm0.017$  & $16.458\pm0.010$  & $-3.137\pm0.107$  & 
$16.480\pm0.032$  & $-3.038\pm0.021$  & $16.463\pm0.012$  & $1.242$   & $0.289$ \\
& $J$~($2.423$)       & $-3.116\pm0.015$  & $16.298\pm0.009$  & $-3.212\pm0.104$  & $16.316\pm0.030$  & $-3.124\pm0.018$    & $16.304\pm0.011$  & 
$1.220$   & $0.295$ \\
& $Ks$~($2.529$)      & $-3.027\pm0.018$  & $16.538\pm0.010$  & $-3.159\pm0.100$  & $16.565\pm0.032$  & $-3.042\pm0.022$    & $16.548\pm0.014$  &
$2.239$   & $0.106$ \\
& $G$~($2.642$)       & $-2.845\pm0.022$  & $17.152\pm0.013$  & $-2.793\pm0.108$  & $17.122\pm0.036$  & $-2.878\pm0.028$    & $17.174\pm0.017$  & 
$1.702$   & $0.183$ \\ 
& $G_{RP}$~($2.471$)  & $-2.978\pm0.019$  & $16.731\pm0.011$  & $-3.215\pm0.111$  & $16.791\pm0.034$  & $-2.980\pm0.023$    & $16.733\pm0.014$  & 
$2.726$   & $0.065$\\ \hline
FO	
& $V$~($2.390$)       & $-3.291\pm0.025$  & $16.739\pm0.008$  
& $-3.394\pm0.039$  & $16.758\pm0.009$  & $-2.695\pm0.088$
& $16.439\pm0.045$  & $21.704$  & \textbf{0.000}\\ 
& $I$~($2.460$)       & $-3.349\pm0.018$  & $16.203\pm0.006$  
& $-3.431\pm0.028$  & $16.217\pm0.007$  & $-2.933\pm0.066$ 
& $15.994\pm0.034$ & $19.444$   & \textbf{0.000} \\
& $G$~($2.506$)       & $-3.281\pm0.028$  & $16.656\pm0.010$  
& $-3.379\pm0.042$  & $16.674\pm0.011$  & $-2.650\pm0.107$ 
& $16.343\pm0.055$  & $15.912$  & \textbf{0.000} \\
& $G_{RP}$~($2.433$)  & $-3.369\pm0.023$  & $16.256\pm0.008$
& $-3.494\pm0.035$  & $16.278\pm0.009$  & $-2.859\pm0.088$
& $16.007\pm0.045$  & $21.981$  & \textbf{0.000} \\
& $Y$~($2.506$)       & $-3.358\pm0.017$  & $15.990\pm0.006$
& $-3.444\pm0.026$  & $16.005\pm0.006$  & $-2.935\pm0.070$
& $15.776\pm0.037$  & $22.564$  & \textbf{0.000} \\
& $J$~($2.460$)       & $-3.394\pm0.015$  & $15.824\pm0.005$
& $-3.488\pm0.023$  & $15.840\pm0.005$  & $-2.991\pm0.060$
& $15.623\pm0.033$  & $30.290$  & \textbf{0.000}  \\ 
& $Ks$~($2.488$)      & $-3.452\pm0.011$  & $15.543\pm0.004$
& $-3.511\pm0.016$  & $15.553\pm0.004$  & $-3.182\pm0.043$
& $15.409\pm0.023$  & $24.442$  & \textbf{0.000} \\  \hline
\multicolumn{10}{c}{SMC}\\ \hline
FU  
& $V$~($2.249$)       & $-3.016\pm0.022$  & $17.903\pm0.009$  & 
$-2.977\pm0.071$  & $17.907\pm0.015$ & $-2.752\pm0.061$
& $17.728\pm0.039$  & $19.752$  & \textbf{0.000} \\
& $I$~($2.421$)       & $-3.199\pm0.018$  & $17.339\pm0.007$  & 
$-3.200\pm0.054$  & $17.348\pm0.012$ & $-2.921\pm0.048$
& $17.154\pm0.027$  & $24.805$  & \textbf{0.000} \\
& $G$~($2.381$)       & $-3.013\pm0.022$  & $17.273\pm0.010$  & $-2.555\pm0.0783$  & 
$17.184\pm0.019$  & $-2.886\pm 0.051$   & $17.184\pm0.019$  & 
$21.872$  & \textbf{0.000} \\
& $G_{RP}$~($2.527$)  & $-3.012\pm0.022$  & $17.302\pm0.010$  & $-2.727\pm0.072$  & 
$17.246\pm0.018$  & $-2.852\pm0.055$    & $17.184\pm0.035$  &
$15.998$  & \textbf{0.000} \\
& $Y$~($2.506$)       & $-3.290\pm0.019$  & $17.196\pm0.008$  & $-3.323\pm0.058$  & $17.210\pm0.013$  & 
$-3.021\pm0.050$  & $17.017\pm0.031$  & $17.715$  & \textbf{0.000} \\
& $J$~($2.654$)       & $-3.307\pm0.018$  & $16.922\pm0.007$  & $-3.413\pm0.050$  & 
$16.948\pm0.012$  & $-3.060\pm0.048$    & $16.759\pm0.031$  & 
$15.395$  & \textbf{0.000} \\
& $Ks$~($2.404$)      & $-3.439\pm0.016$  & $16.669\pm0.006$  & $-3.505\pm0.051$  & 
$16.689\pm0.011$  & $-3.211\pm0.037$    & $16.520\pm 0.023$ & 
$21.439$  & \textbf{0.000} \\ \hline
FO  
& $V$~($2.576$)       & $-3.166\pm0.033$  & $17.180\pm0.008$
& $-3.267\pm0.042$  & $17.181\pm0.008$  & $-2.563\pm0.353$ 
& $16.944\pm0.176$  & $8.788$   & \textbf{0.000}\\
& $I$~($2.471$)       & $-3.321\pm0.027$  & $16.696\pm0.006$
& $-3.392\pm0.035$  & $16.696\pm0.006$  & $-2.665\pm0.247$
& $16.411\pm0.120$  & $7.167$   & \textbf{0.000}\\ 
& $G$~($2.570$)       & $-3.196\pm0.036$  & $17.119\pm0.009$
& $-3.290\pm0.047$  & $17.122\pm0.009$  & $-2.559\pm0.350$
& $16.852\pm0.173$  & $6.038$   & \textbf{0.000} \\
& $G_{RP}$~($2.624$)  & $-3.297\pm0.033$  & $16.745\pm0.008$  
& $-3.375\pm0.043$  & $16.748\pm0.008$  & $-2.790\pm0.330$
& $16.535\pm0.165$  & $5.027$   & \textbf{0.000} \\
& $Y$~($2.750$)       & $-3.352\pm0.027$  & $16.570\pm0.006$
& $-3.423\pm0.034$  & $16.572\pm0.006$  & $-2.749\pm0.375$ 
& $16.312\pm0.192$  & $7.124$   & \textbf{0.000} \\
& $J$~($2.630$)       & $-3.412\pm0.025$  & $16.350\pm0.006$
& $-3.473\pm0.033$  & $16.352\pm0.006$  & $-2.908\pm0.282$
& $16.131\pm0.142$  & $5.878$   & \textbf{0.000}  \\
& $Ks$~($2.618$)      & $-3.539\pm0.022$  & $16.123\pm0.005$  
& $-3.587\pm0.029$  & $16.124\pm0.005$  & $-3.073\pm0.220$
& $15.915\pm0.110$  & $4.818$   & \textbf{0.000}  \\ \hline \hline
\end{tabular}
\label{tab:slopes_pl_maxmin_lmcsmc}
\end{center}
\end{table*}

\subsubsection{Colour Magnitude Diagram (CMD)}
\begin{figure}
\begin{tabular}{c}
\resizebox{1.0\linewidth}{!}{\includegraphics*{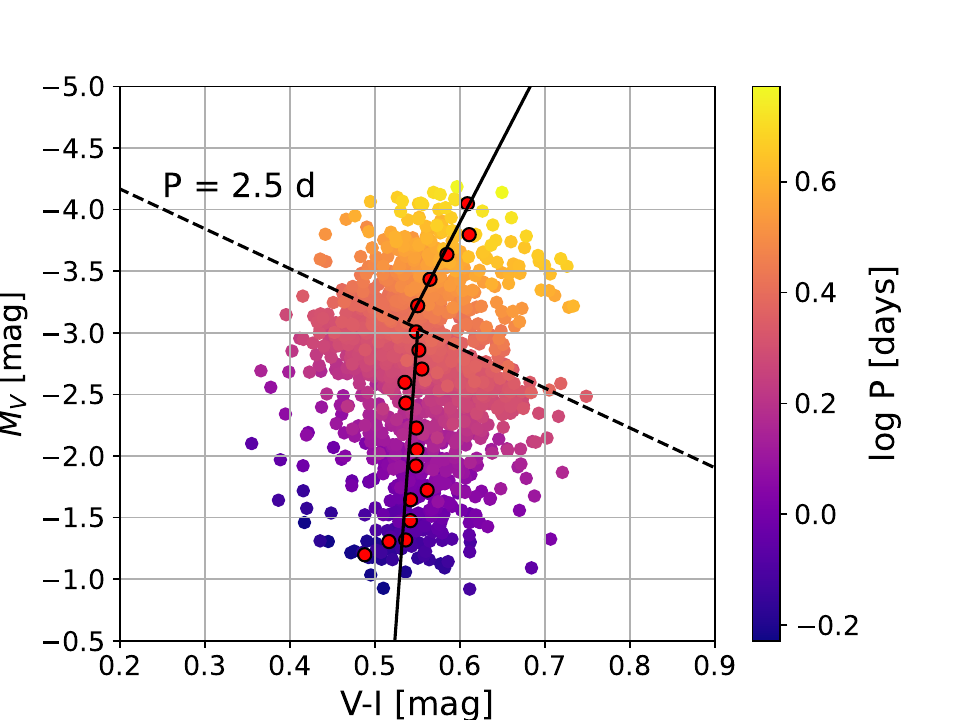}}
\end{tabular}
\caption{CMD of LMC FO Cepheids for $V$ and $I$ bands. The FO Cepheids samples are binned according to their periods, and the mean color indices and absolute magnitudes within each bin are computed and plotted on the CMD. These mean values are represented by red circles. Solid black lines represents the linear fits for stars with periods shorter and longer than $P=2.5$ d. A dashed line marks the constant period at $P = 2.5$ d, while the colour bar denotes the period associated with each individual Cepheids. }
\label{fig:cmd_lmc_vi_obs}
\end{figure}

We now investigate the break-point in the CMD using FU/FO Cepheids for the LMC and SMC. It is well known 
that the PC/PL/AC relations are closely related to each other; a change in the PC relation has a direct impact 
on the PL relation. This motivates us to extend to the CMD plane. Fig.~\ref{fig:cmd_lmcsmc_fufo_obsmodels} displays the CMD using $(V, I)$-bands for both observations and models. For both LMC and SMC, it can be observed from the figure that the FO models using convection sets B and D lie towards the higher $(V-I)$ end of the CMD. The CMD constructed from the theoretical models using convection sets B and D for SMC lies around a mean of $(V-I)=$ $0.529$ mag and $0.578$ mag, respectively. Similarly, for the LMC using the same convection sets, the CMD is found to lie around a mean of $(V-I)=$ $0.565$ mag and $0.627$ mag, respectively. 

For LMC FO Cepheids, only stars with  $P> 0.58$~d were selected to examine the potential break-point in the CMD at 
$P=2.5$~d. A corresponding analysis was also performed for FU Cepheids in both the LMC and SMC at $P=2.5$ d. In each case, the Cepheids samples were binned according to period, and the mean color index and absolute magnitude were computed for each bin. The plot is shown in Fig.~\ref{fig:cmd_lmc_vi_obs}. The statistical significance of the CMD break-point at $P= 2.5$~d was evaluated using the $F$- test for FO Cepheids in both galaxies, with the same test applied to the FU Cepheids samples in the LMC and SMC. Table \ref{tab:slopes_cmd_lmcsmc} summarised the coefficients of the CMD and the results of the $F$- test at mean light for both LMC/SMC FO and FU Cepheids. 

The break-point at $P=2.5$~ d in the CMD is an approximate estimate intended to highlight its potential connection with the break-point observed in the PC, PL, and AC relations. The break-point in the PC/PL/AC relations translates into the break-point in the CMD. However, a more detailed and rigorous investigation is necessary to confirm and fully characterize this connection. Recently, \citet{espi24} investigated the empirical positions of the IS of the classical Cepheids in the LMC. They also investigated the break-point in the IS edges at $P=3$ d. The authors argued that the occurrence of the break-point at $P=3$ d is caused by the depopulation of second- and third-crossing Cepheids in the faint part of the IS.

\begin{table*}
\begin{center}
\caption{Results of $F$-test for CMD for FO Cepheids in the LMC/SMC to check the break-point at $P= 2.5$ d. Same has been conducted for FU Cepheids to investigate the break-point at $P=2.5$ d. The bold-faced entries indicate the break-point in the CMD.}
\begin{tabular}{c c c c c c c c c c c c c} \\ \hline \hline
Type&   &  $a_{\rm all}$ & $b_{\rm all}$ & $a_{\rm s}$ & $b_{\rm s}$ & $a_{\rm l}$ & $b_{\rm l}$ & $F$ & $p(F)$     \\ \hline 
	\multicolumn{9}{c}{LMC}\\ \hline
FO
& $V-I$     &   $-25.238$   &   $11.409$    &   $-93.333$   &   $48.333$    &   $-13.333$   &   $4.100$     
&   $12.671$    &   $\textbf{0.000}$   \\
& $V-Y$     &   $-18.928$   &   $11.914$    &   $-26.363$   &   $17.200$    &   $-7.580$    &   $2.459$     
&   $518.022$   &   $\textbf{0.000}$   \\
& $V-J$     &   $-19.629$   &   $11.774$    &   $-26.250$   &   $16.325$    &   $-7.833$    &   $2.333$     
&   $520.321$   &   $\textbf{0.000}$  \\
& $V-Ks$    &   $-11.777$   &   $12.166$    &   $-11.666$   &   $12.166$    &   $-5.465$    &   $3.626$     
&   $499.899$   &   $\textbf{0.000}$ \\
& $V-G$     &   $-45.833$   &   $0.916$     &   $-60.000$   &   $1.800$     &   $-29.411$   &   $-1.088$    
&   $232.452$   &   $\textbf{0.000}$ \\
&$V-G_{RP}$ &   $-22.000$   &   $8.580$     &   $-23.076$   &   $9.2300$    &   $-9.400$    &   $1.462$     
&   $332.593$   &   $\textbf{0.000}$ \\
\hline
FU 
&   $V-I$   &   $-13.157$   &   $5.789$     & $-13.529$   &   $6.088$       &   $-12.500$     &   $5.350$     
&   $297.360$   &   $\textbf{0.000}$  \\
&   $V-Y$   &   $-9.636$    &   $5.870$     & $-7.777$    &   $4.388$       &   $-8.888$      &   $5.166$     
&   $285.819$   &   $\textbf{0.000}$ \\
&   $V-J$   &   $-6.235$    &   $4.164$     & $-5.416$    &   $3.495$       &   $-7.560$      &   $5.690$     
&   $284.740$   &   $\textbf{0.000}$ \\
&   $V-Ks$  &   $-5.520$    &   $5.320$     & $-3.620$    &   $2.877$       &   $-5.961$      &   $6.005$     
&   $257.877$   &   $\textbf{0.000}$ \\
&   $V-G$   &   $-34.375$   &   $1.031$     & $25.555$    &   $-4.088$      &   $-34.444$     &   $1.044$     
&   $288.482$   &   $\textbf{0.000}$      \\
&$V-G_{RP}$ &   $-13.414$   &   $5.231$     & $-13.529$   &   $5.276$       &   $-15.500$     &   $6.590$     
&   $288.299$   &   $\textbf{0.000}$   \\ 
\hline
\multicolumn{9}{c}{SMC}\\ \hline
FO 
& $V-I$     &   $-18.518$   &   $7.592$     &   $-25.000$   &   $10.750$    &   $-6.000$    &   $0.300$    
&   $32.045$    &   $\textbf{0.000}$ \\ 
& $V-Y$     &   $-20.000$   &   $12.000$    &   $-26.666$   &   $15.666$    &   $-4.423$    &   $0.128$    
&   $214.424$   &\textbf{0.000} \\
& $V-J$     &   $-11.320$   &   $8.245$     &   $-18.181$   &   $14.090$    &   $-3.709$    &   $0.535$    
&   $208.770$   &   $\textbf{0.000}$ \\
& $V-Ks$    &   $-7.792$    &   $7.077$     &   $-13.114$   &   $12.803$    &   $-3.066$    &   $0.733$     
&   $208.989$   &   $\textbf{0.000}$   \\
& $V-G$     &   $-54.545$   &   $0.454$     &   $-68.965$   &   $0.793$     &   $-30.000$   &   $-1.200$    
&   $130.552$   &   $\textbf{0.000}$ \\
& $V-G_{RP}$&   $-13.953$   &   $4.069$     &   $-18.181$   &   $5.727$     &   $-8.888$    &   $1.166$     
&   $123.580$   &   $\textbf{0.000}$ \\
\hline
FU 
&   $V-I$   &   $-15.625$   &   $7.656$     &   $-16.428$   &   $8.214$     &   $-12.000$   &   $5.160$     
&   $123.876$   &   $\textbf{0.000}$ \\
&   $V-Y$   &   $-11.320$   &   $7.226$     &   $-12.777$   &   $8.305$     &   $-7.750$    &   $3.900$     
&   $158.020$   &   $\textbf{0.000}$ \\
&   $V-J$   &   $-11.538$   &   $10.461$    &   $-9.200$    &   $8.004$     &   $-9.393$    &   $7.845$    
&   $153.295$   &   $\textbf{0.000}$    \\
&   $V-Ks$  &   $-7.317$    &   $8.170$     &   $-5.333$    &   $5.333$     &   $-5.263$    &   $4.915$    
&   $156.318$   &   $\textbf{0.000}$   \\
&   $V-G$   &   $-36.363$   &   $0.636$     &   $-30.000$   &   $0.400$     &   $-31.000$   &   $0.180$     
&   $436.315$   &   $\textbf{0.000}$    \\
&$V-G_{RP}$ &   $-13.953$   &   $5.465$     &   $-15.000$   &   $6.250$     &   $-12.307$   &   $4.346$     
&   $425.269$   &   $\textbf{0.000}$ \\
\hline \hline
\end{tabular}
\label{tab:slopes_cmd_lmcsmc}
\end{center}
\end{table*}

\subsection{Empirical PC/PL/AC relations for LMC FO Cepheids with $P<2.5$ d}
Recently, using $YJKs$ bands, \citet{ripe22} reported a break-point in the PL relations of FO Cepheids in the LMC at 
$P = 0.58$~d for which they attribute the break-point may be due to the passage of DCEPS from the ﬁrst crossing to the second. To support this hypothesis, the authors determined a ratio of 8 when comparing the time taken by a $5M{\odot}$ star to cross the IS during its second crossing in the opposite direction to that of a $3M{\odot}$ star during its first crossing. Additionaly, the authors also found a ratio of 4 when calculating the number of the second crossing DCEPS to those in their first. These findings suggest that approximately 4500 DCEPs with $P > 0.58$ d are in their second or third crossing, in contrast to only $140$ stars with $< 0.58$ d, supporting the proposed hypothesis. Piecewise regression analysis also detected the occurence of a break-point around $P=0.58$~d in the both the PC and PL planes. To test the significance of the break-point in the PC/PL relations near $P=0.58$~d, we have selected only the stars with $P<2.5$~d. The results of the
$F$-test are summarised in Tables \ref{tab:fvalue_pc_lmcsmc} and \ref{tab:fvalue_pl_lmcsmc} in Appendix \ref{sec:ftest_pw}. The statistical $F-$ test shows the evidence of break-points near $P=0.58$~d in both the PC and PL planes. We have excluded the Gaia bands because of the small numbers of light curves with $P<0.58$ ($<20$ number of stars). 
We also found that the break-point in the PC relations at maximum, minimum is significant for all colour indices ($V-I, V-Y, V-J, V-Ks$). The same is true for the AC relations. The break-point in the PL relations is found to be significant for all photometric bands ($VIYJKs$) at maximum, minimum light. The coefficients of the empirical PC/PL/AC relations are summarised in the supplementary data.

\subsection{Cepheids evolutionary models}
Using stellar evolutionary tracks for Cepheids in the mass range between $2 - 6 M_{\odot}$ for both LMC and SMC composition, we attempt to understand the nature of the break-point at a particular period. Such similar study is carried out by \citet{ripe22} to understand the origin of the break-point in the PL relations of the FO Cepheids in the LMC at $P=0.58$ d.

Recently, \citet{Ziol24} computed a comprehensive grid of Cepheids models with \textsc{MESA} varying the 
choices of the input physics (opacity tables, scaled solar abundances, atomic diffusion, nuclear reaction network, atmosphere tables, MLT and convective theories,  numerical convergence, convective core overshooting). The choices for atmosphere model, reference solar composition, nuclear reaction rates, and scheme to determine convective boundaries can have a much stronger impact on evolutionary tracks. This impact is sometimes barely noticeable, but for some of the options, significant shifts and even qualitative differences in evolutionary tracks were recorded. We computed stellar evolutionary tracks for Cepheids in the mass range between $2-6~M\odot$ for both LMC and SMC composition. However, this study provides no solid physical justification behind a specific option. We utilized the 'inlist' provided in $5M\_Cepheids\_blue\_loop$ in the $test\_suite$ of \textsc{mesa}-r15140 with a few modifications of the choices that largely impact the shape of the blue loops of the evolutionary tracks computed. The choices have been made after testing a few choices of the input parameters in reference to the work done by \citet{Ziol24} on the extension of the blue loop entering the IS. The evolutionary tracks are computed from the pre-main sequence till the completion of the blue loop. The 'inlist' ($\rm inlist\_rsp\_cmd$) used in computing the evolutionary track is available at \url{https://github.com/Kerdaris/FO_MC_CMD}. 

\begin{figure*}
\begin{tabular}{cc}
\resizebox{1.0\linewidth}{!}{\includegraphics*{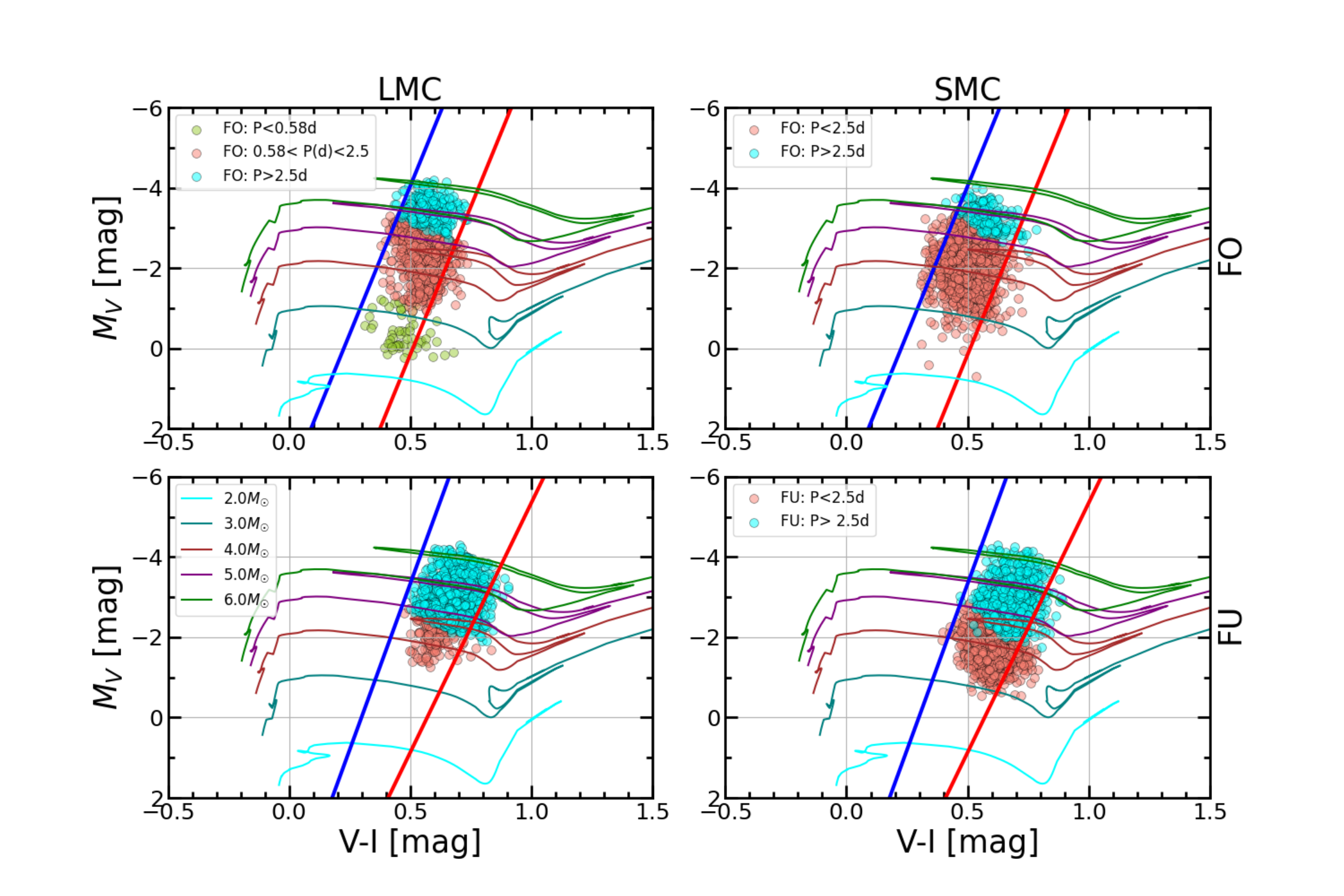}}& 
\end{tabular}
\caption{Evolutionary tracks of Cepheids shown from zero age main sequence to core He exhaustion composition for 
LMC (left panel) and SMC (right panel) in the CMD plane. The tracks are computed for masses $(2-6)~M_{\odot}$ 
in steps of $1~M_{\odot}$. The tracks are overplotted with the CMD of FU (lower panel) and FO (upper panel) Cepheids, respectively. The IS edges are taken from \citet{deka24} where blue and red represents the blue and red edges of the IS. }
\label{fig:etrack_cmd_lmcsmc_fufo_obs}
\end{figure*}

Fig.~\ref{fig:etrack_cmd_lmcsmc_fufo_obs} represents the evolutionary tracks of Cepheids models for LMC/SMC 
composition in the CMD plane. The CMD for FU/FO Cepheids using OGLE-IV data are overplotted in the tracks 
displayed in Fig.~\ref{fig:etrack_cmd_lmcsmc_fufo_obs}. For this particular computation adopted in this study for both LMC and SMC, it can be seen from Fig.~\ref{fig:etrack_cmd_lmcsmc_fufo_obs} that for stars with $M<4M\odot$, the blue loop does not enter into the IS. However, the precise value of the stellar masses that do not enter the IS depends on the input physics adopted. For FO Cepheids in the LMC, it is observed that for $M<4M\odot$ the instability strip is populated by stars on the first crossing. These are mainly stars with $P<0.58$~d and are likely stars in the sub-giant branch. This is in agreement with the results reported in \citet{ripe22}. Similarly, the evolutionary tracks computed by \citet{espi24} also depicts similar results for $M<3.7M\odot$. In both cases, they found that the blue loop of the stars whose stellar mass is approximately $<4.0M\odot$ does not extend blueward enough to cross the IS. On the other hand, it is seen that for $M>4M\odot$ the instability strip is populated by stars in the second and third crossing. It is important to note that stars in the second/third crossings are at a higher luminosity than the ones in the first crossing. Furthermore, FO Cepheids with $0.58<P(d)<2.5$ in both LMC and SMC are a mixture of stars in the sub-giant branch and blue loop. Most of the FU/FO Cepheids with $P>2.5$ d for LMC/SMC are likely in the blue loop. 

It can be seen from Fig.~\ref{fig:etrack_cmd_lmcsmc_fufo_obs} that for FO Cepheids in the LMC, the width of the CMD in the period range $0.58<P (d) <2.5$ is narrow. The corresponding PC relations for this period range was found to be flat, indicating that the stars appear to have nearly the same colour despite their varying magnitudes and periods. These claims necessitate further detailed and in-depth investigation in a separate study to understand the mechanism and the physics behind the break-point at $P\sim2.5$ d and $P\sim 0.58$ d.

\section{Summary and Conclusion}
\label{sec:conclusion}
In this study, we present a detailed light curve analysis of FO/FU Cepheids in the LMC and SMC using both observed data and theoretical models. The observed data are obtained from OGLE-IV, Gaia-DR3, and VMC databases. The theoretical light curves are obtained using version r15410 of \textsc{mesa-rsp}. The light curves are studied in terms of the Fourier parameters as a function of period, PA relation, CMD at mean light and in terms of PC/PL/AC relations at maximum, minimum, and mean light. We thoroughly investigated the occurence of break-points in the empirical PC and PL relations for different colour indices and photometric bands for the LMC and SMC FU/FO Cepheids using piecewise regression analysis and statistical $F$- test. The results and conclusion are summarised below:

\begin{enumerate}
\item The PA relation of the FO Cepheids in the LMC and SMC display interesting results. For shorter periods 
($P<2.5$~d), the SMC Cepheids are found to have higher amplitudes than those in the LMC within the same 
period range. This is consistent with the findings of \citet{klag09,szab12}. 
\item The Fourier parameters of FO Cepheids as a function of period display sharp changes around $P=2.5$~d in $V$  and $I$-bands. This variation in the FPs is also seen in $I$-band \citep{bhar14}. The same is also found in $G$ band. The results obtained from the theoretical models also exhibit similar results. When compared with 
observations, the FPs obtained using Set D are found to be more consistent than the set B for both LMC and SMC. 
Furthermore, we also observed a variation in the FPs of the LMC FO Cepheids near $P=0.58$~d. The change in the FPs around $P=0.58$~d and $P=2.5$~d may be closely related to the occurrence of the break-point in the PC/PL and AC relations at these periods.
\item Using piecewise regression, we observed multiple break-points in the PC/PL relations for the LMC and SMC FO/FU Cepheids. The location of the break-points varies across different colour indices and photometric bands. The break-points obtained from the piecewise regression analysis are tested using the statistical $F$-test and are found to show different levels of significance. Some of these break-points are found to have higher confidence levels as compared to the break-points reported in the literature. Although some of the break-points obtained are found to be significant, some are not. 
\item Some of the break-points obtained using piecewise regression are found to be commonly observed in the PC and PL planes across all colour indices and photometric bands. The break-points near $P=2.5$~d and $P=4.0$~d are detected in the case of PC and PL relations using LMC FU Cepheids. Although the break-points near $P=2.5$~d and $P=4.0$~d are not significant in the PC plane, some of these break-points are found to be significant in the PL plane. On the other hand, the break-points near $P=2.5$~d are seen in the PC and PL relations using SMC FU Cepheids. Some of these break-points are found to be significant in both the PC and PL planes whereas some are not. For both LMC and SMC FO Cepheids, the break-points around $P=1.0$~d and $P=2.5$~d have been observed in the PC and PL relations. These break-points are found to be significant in both the PC and PL planes in the case of LMC FO Cepheids. In addition, the break-points near $P=0.58$~d using $VIYJKs$ photometric bands are also found to be significant in both PC and PL planes. For the SMC FO Cepheids, the break-points near $P=1.0$~d and $P=2.5$~d are found to be significant in the PC plane, whereas, only the break-points near $P=2.5$~d are found to be significant in the PL plane. 
\item We found that for the SMC FO Cepheids, the slope of the PC relations increases for shorter periods ($\sim P<2.5$~d) but the increase in the slope is more for longer periods ($\sim P>2.5$~d). However for the LMC, the slope of the PC relation is steeper only for Cepheids having periods $P<0.58$~d and $P>2.5$~d in all the colour indices. It is interesting to note that the slope of the PC relations for LMC FO Cepheids is found to be shallow for Cepheids within the period range $0.58< P(d)<2.5$ obtained using $V-I, V-G$ and $V-G_{RP}$. However, we found that the slope is increasing for $V-Y, V-J$ and $V-Ks$ in the same period range. Overall, the slope of the PC relations is steeper for Cepheids having periods $P<0.58$~d and $P>2.5$~d in all the colour indices. 
\item The flat slope in the PC relations of LMC FO Cepheids for $0.58<P(d)<2.5$ in $V-I, V-G$, and $V-G_{RP}$ is explained using the CMD. We found that the width of the CMD is narrow for this period range. Hence, stars within this period range seem to have nearly the same colour but different magnitudes and periods. This is one of the important results of this study.
\item Using piecewise regression analysis on the theoretical data points, we found break-points near $P=2.5$~d in the PC/PL relations for the LMC and SMC FO/FU Cepheids using convection sets B and D across all colour indices and photometric bands. The detected break-points are tested using statistical $F$- test and are found to be significant.  
\item The slopes of the theoretical PL relations are steeper with the increase in wavelength. The dispersions of the PL relation decrease significantly from the optical to the near-infrared passbands for both LMC and SMC. When compared with observations, the relations obtained using convection set B is found to be more consistent than the convection set D.
\item One of the important results of this study is the correlation of the PC/PL/AC relations with the
CMD. We found that the break-point near $P=2.5$~d in the PC and PL relations for the LMC and SMC FO Cepheids is also evident in the AC and CMD planes. A similar trend is also observed for the LMC and SMC FU Cepheids, where the break-point near $P=2.5$~d  in the PC and PL planes is reflected in both the AC and CMD planes.
\item For FO Cepheids in the LMC, we found a break-point in the empirical PL relations near $P=0.58$~d in $VIYJKs$ photometric bands at mean light. The same is true for maximum and minimum light. These results support the finding of \citet{ripe22}. The same break-point was also observed in the empirical PC and AC relations. 

\end{enumerate}

We did not find any break-point in the empirical PC/PL/AC relations of FO Cepheids in the SMC near $P=0.58$~d. This may be due to statistically fewer number of data points below $P<0.58$~d ($\sim 2.18\%$ of $1345$ stars). Due to the same reason, this has been left out for the theoretical models. \citet{mado91} discussed the idea of PLC plane at mean light, where PL and PC relations represent the PLC plane viewed in different ways. However, this work 
finds strong evidence that this can be extended to all pulsation phases and can be used to explore the different envelope physics at these different phases. Recently, \citet{deka24} has constructed a large grid of linear models of FO Cepheids covering the lowest period range using Sobol numbers both for LMC ($P=0.249$~d) and SMC ($P=0.252$~d) composition detected 
by observations. The need for a large grid of non-linear FO models that covers the whole range of observational
data is crucial in understanding the pulsational physics of FO Cepheids. We would like to point out that the non-linear models computed in this study using the four sets of convection parameters outlined in \citet{paxt19} are 
preliminary. The theoretical PC/PL/AC relations derived using the models in the present study should be taken as  preliminary due to the limitations of the current grid. An alternative approach by constructing the model grid using Sobol numbers that are uniformly distributed and more rapidly fill the input parameter space may enhance the understanding behind the potential break-points in the theoretical PC, PL and AC relations. Moreover, a detailed investigation regarding the discrepancy in the Fourier amplitudes and the bumps in the theoretical light curves of FO Cepheids will shed more light on the role of turbulent convection in stellar 
pulsation. Such studies will also provide further constraints in pulsation models. These investigations hold significant importance in understanding the underlying physics of FO Cepheids and is a subject of future studies.  

\section*{Acknowledgements}
\addcontentsline{toc}{section}{Acknowledgements}
The authors are grateful to the anonymous referee for carefully reading the manuscript and providing valuable comments and helpful suggestions which have significantly improved its presentation. KK thanks the Council of Scientific and Industrial Research (CSIR), Govt. of India for the Senior Research Fellowship (SRF). This research was supported by the International Space Science Institute (ISSI) in Bern/
Beijing through ISSI/ISSI-BJ International Team project $ID \#24-603$ - “EXPANDING Universe” (EXploiting 
Precision AstroNomical Distance INdicators in the Gaia Universe). SMK thanks support from SUNY Oswego. SD thanks 
Council of Scientific and Industrial Research (CSIR), Govt. of India, New Delhi for a financial support through the
research grant ``03(1425)/18/EMR-II''. MD acknowledges the funding from the Large grant INAF 2023 MOVIE (PI: M. Marconi). S.D. acknowledges the KKP-137523 `SeismoLab' \'Elvonal grant of the 
Hungarian Research, Development and Innovation Office (NKFIH). GB is grateful to the Department of Science and 
Technology (DST), Govt. of India, New Delhi for providing financial support through the DST INSPIRE Fellowship 
research grant (DST/INSPIRE/Fellowship/2019/IF190616). The authors acknowledge the use of highly valuable publicly 
accessible archival data from OGLE-IV, Gaia DR3, and VMC surveys and IUCAA, Pune for the use of the High-
Performance Computing facility Pegasus. The authors also acknowledge the use of MESA-r15140 software in  this 
project \citep{paxt11, paxt13, paxt15, paxt18, paxt19}. 

\section*{Data Availability}
The \citet{skow21} map is available from \url{http://ogle.astrouw.edu.pl/cgi-ogle/get_ms_ext.py}. The OGLE-IV 
data is downloaded from \url{http://ftp.astrouw.edu.pl/ogle/ogle4/OCVS/}. Using Python astro-query, the Gaia 
DR3 data is downloaded from \url{http://cdn.gea.esac.esa.int/Gaia/gdr3/}. The near-infrared data for the LMC 
and SMC are downloaded from \url{https://academic.oup.com/mnras/article/512/1/563/6544651} and 
\url{https://vizier.cds.unistra.fr/viz-bin/VizieR?-source=J/ApJS/224/21}. The theoretical models will be made 
available on reasonable request to the corresponding authors. The data underlying this article are available in its online supplementary material.

\bibliographystyle{mnras}
\bibliography{mc_fufo}

\appendix
\clearpage

\section{HR Diagram}
\label{sec:hrd}
Hertzsprung Russel Diagram (HRD) for LMC/SMC FO Cepheids models obtained using Set B and Set D. Blue, red and brown represents IS edge from \citet{somm22}, \citet{deka24} and \citet{espi24}, respectively.   
\begin{figure*}
\begin{tabular}{cc}
\resizebox{1.0\linewidth}{!}{\includegraphics*{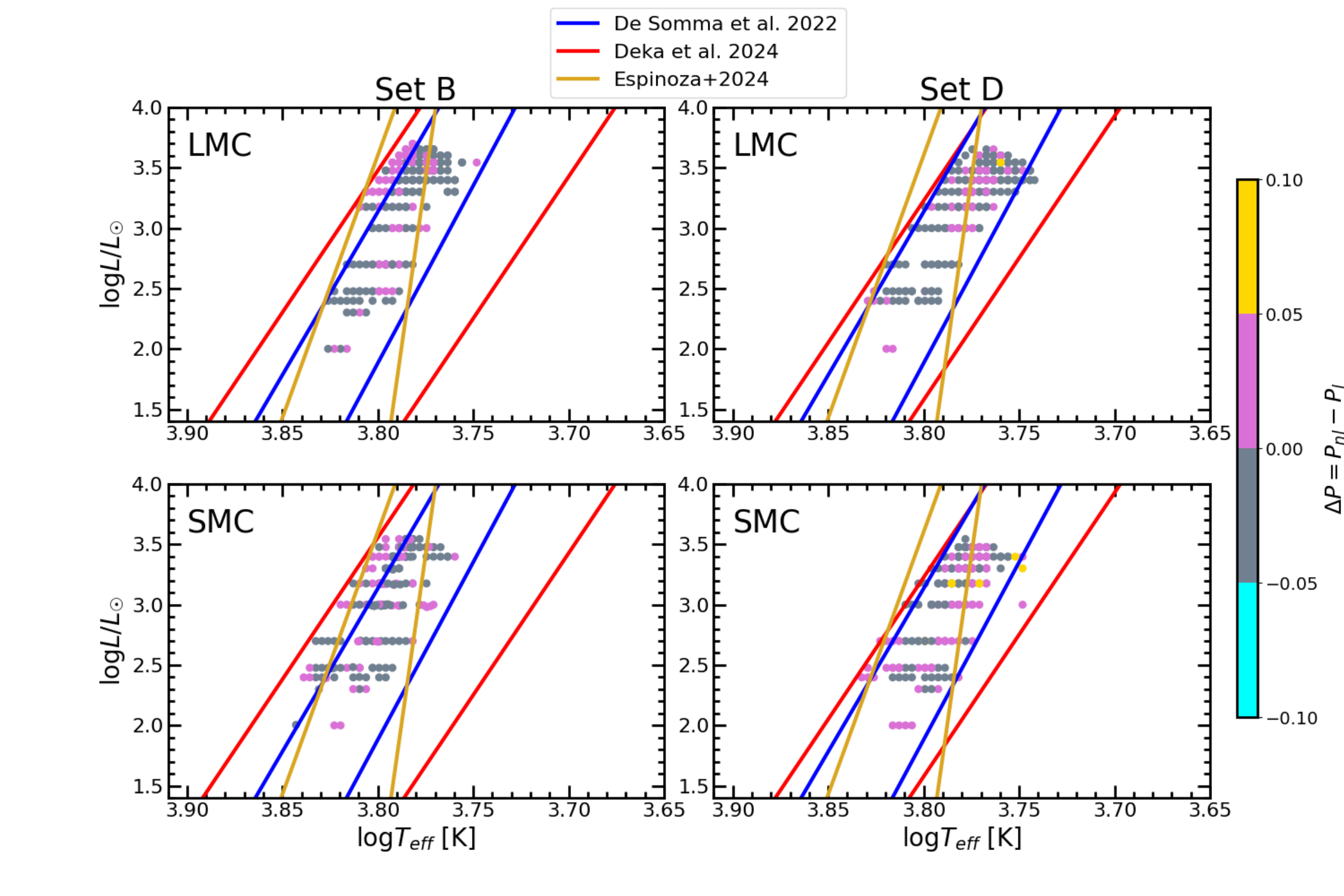}}& 
\end{tabular}
\caption{Hertzsprung Russel Diagram (HRD) for the LMC and SMC FO Cepheids models obtained using Set B and Set D. Blue, red and brown represents IS edge from \citet{somm22}, \citet{deka24} and \citet{espi24}, respectively. The colourbar shows the difference between the non-linear period ($P_{nl}$) and linear period ($P_{l}$) of the models, i.e $\Delta P = P_{nl}-P_{l}$. The individual mean difference between the linear and non-linear periods in our models is $\sim 0.02$ d. }
\label{fig:HR_lmcsmc_bd}
\end{figure*}

\section{PC and PL relations}
\begin{figure}
\begin{tabular}{c}
\resizebox{1.0\linewidth}{!}{\includegraphics*{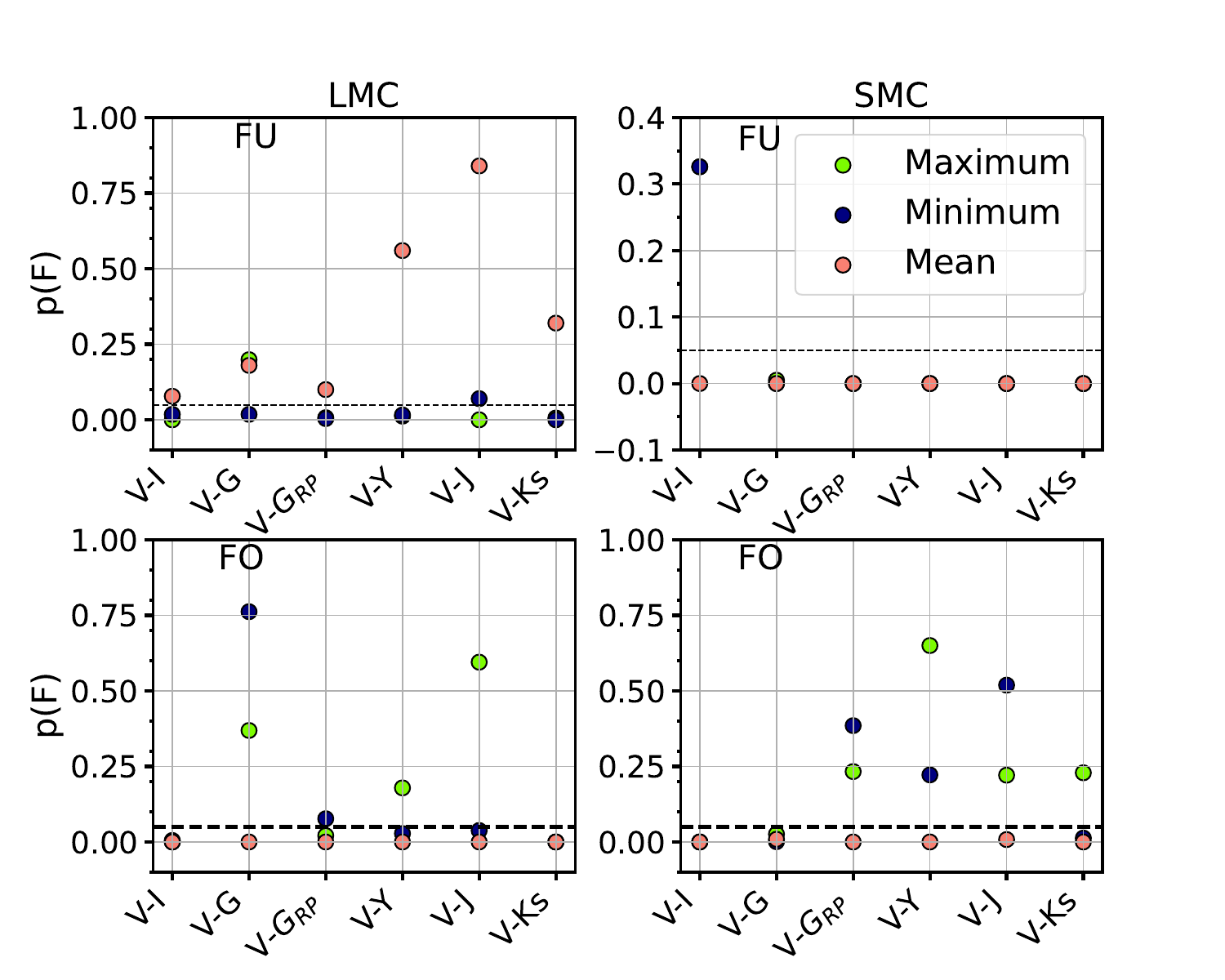}}
\end{tabular}
\caption{Probabilities of the $F$-test of the PC relations for LMC and SMC FO/FU Cepheids. Green, blue, and orange represents maximum, minimum, and mean light, respectively. Black dotted line represents the value of the probability of the $F$-test at $95\%$ confidence, p(F)=0.05.}
\label{fig:ftest_pc}
\end{figure}

\begin{figure}
\begin{tabular}{c}
\resizebox{1.0\linewidth}{!}{\includegraphics*{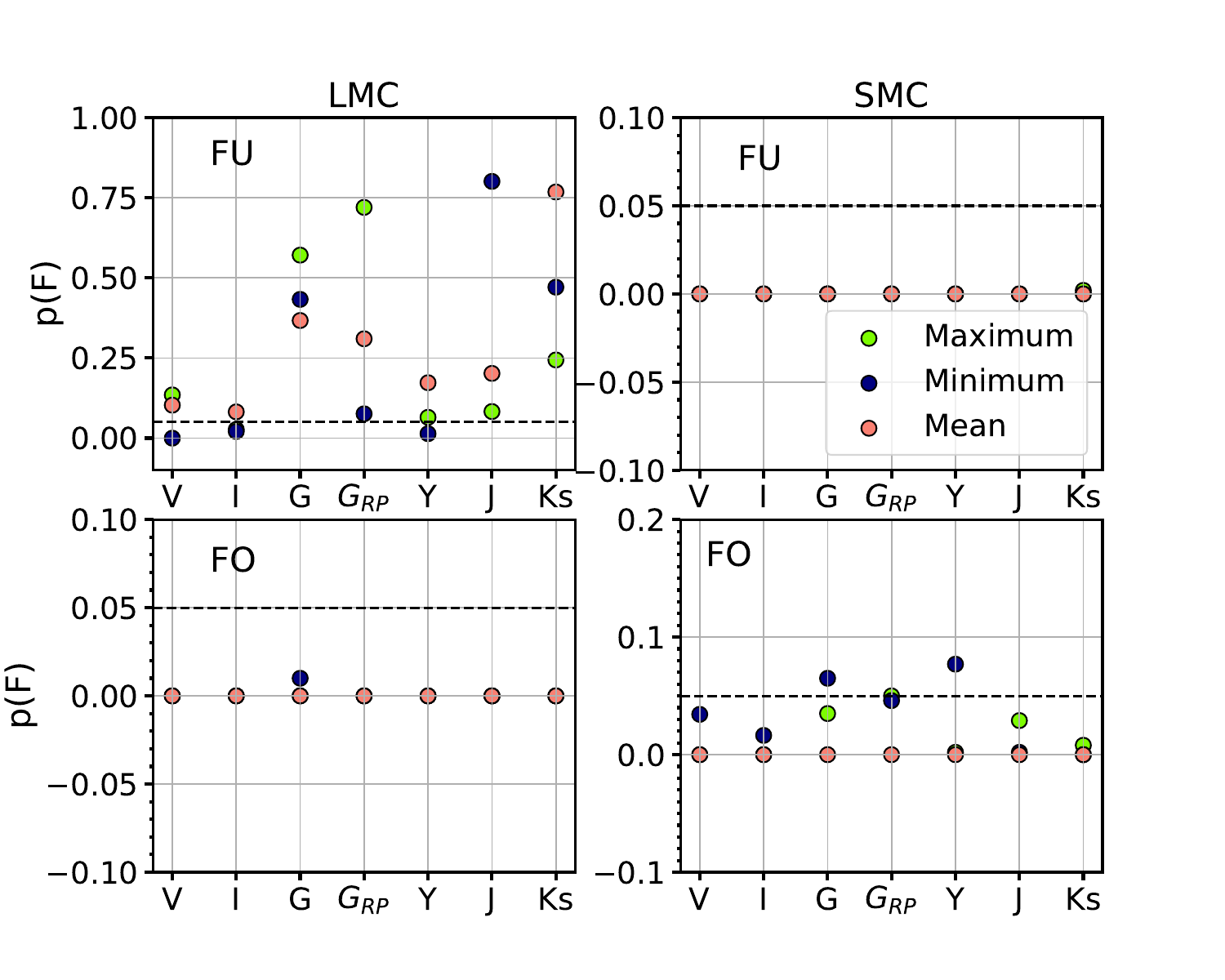}}
\end{tabular}
\caption{Same as Fig.~\ref{fig:ftest_pc} but for PL relations.}
\label{fig:ftest_pl}
\end{figure}

\begin{figure}
\begin{tabular}{cc}
\resizebox{1.0\linewidth}{!}{\includegraphics*{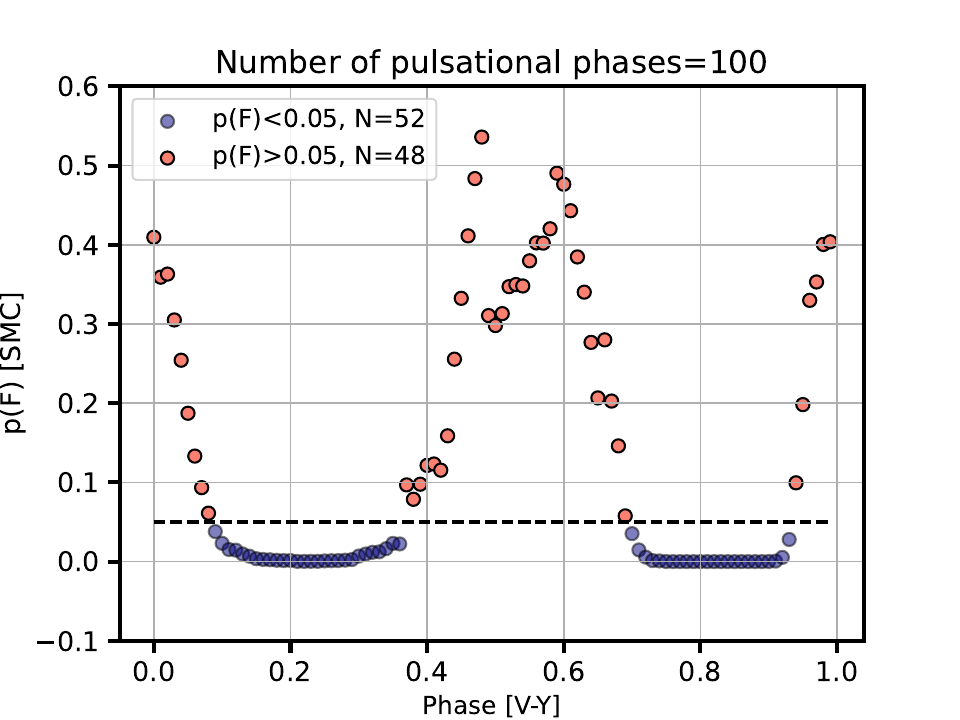}}& 
\end{tabular}
\caption{Probability of the statistical $F$-test of PC relation in $V-Y$ colour for the SMC FO Cepheids at $100$ pulsational phase. The black dotted line represents the $95\%$ confidence level, p(F)=0.05. Orange and purple represent p(F) > 0.05 and p(F) < 0.05, respectively.}
\label{fig:probability_ftest}
\end{figure}

Fig.~\ref{fig:ftest_pc} and ~\ref{fig:ftest_pl} represent the probability of the $F$-test of the PC relations at maximum, minimum, and mean light for FU (upper panel) and FO Cepheids (lower panel) for both LMC (left panel) and SMC (right panel). In Figs.~\ref{fig:ftest_pc} and \ref{fig:ftest_pl}, the green dotted lines represent p(F)=0.05. Fig.~\ref{fig:probability_ftest} represents the probability of the $F$-test for PC relation of FO Cepheids in the SMC in $V-Y$ for $100$ pulsational phase. 

\label{sec:f_test_pcpl}

\section{$F$-test for the PC/PL relations of FU and FO Cepheids in the LMC/SMC obtained using piecewise regression}
\label{sec:ftest_pw}
Tables \ref{tab:fvalue_pc_lmcsmc}, \ref{tab:fvalue_pc_fu_lmcsmc}, \ref{tab:fvalue_pl_lmcsmc}, and \ref{tab:fvalue_pl_fu_lmcsmc} summarised the results of the $F$- test of PC/PL relations for LMC/SMC FU and FO Cepheids at different break-points obtained using the piecewise regression analysis. 

\begin{table*}
\begin{center}
\caption{Results of the $F$-test for break-points in the PC relations using piecewise regression from the observed data at mean light for the LMC and SMC FO Cepheids. The bold-faced entries indicate the significance of the break-points in the PC relations. The first ($P=2.5$ d) and second ($P=0.58$ d) rows indicate the results of the $F$- test for the corresponding break-points as reported in the literature. The remaining rows represent the $F$-test results for break-points obtained using piecewise regression. \textbf{The uncertainties represent standard errors
based on $95\%$ confidence interval obtained from piecewise regression. }}
\begin{tabular}{c c c c c c c} \\ \hline \hline
$V-I$    &    $V-G$   & $V-G_{RP}$  & $V-Y$ & $V-J$   & $V-Ks$   \\ \hline
\multicolumn{6}{c}{LMC}\\ \hline
$P=2.5$~d   &   $P=2.5$~d   &   $P=2.5$~d     &   $P=2.5$~d     &   $P=2.5$~d       & $P=2.5$~d \\
$F=18.367$  &   $F=6.443$   &   $F=8.773$     &   $F=12.495$    &   $F=11.442$      & $F=11.692$ \\
$\textbf{p(F)=0.000}$             & $\textbf{p(F)=0.000}$               &   $\textbf{p(F)=0.000}$             &
$\textbf{p(F)=0.000}$             & $\textbf{p(F)=0.000}$               &   $\textbf{p(F)=0.000}$ \\ \hline
$P=0.58$~d  &     -         &      -          & $P=0.58$~d      &   $P=0.58$~d      &   $P=0.58$~d    \\
$F=9.417$   &     -         &      -          & $F=7.886$       &   $F=7.427$       &   $F=12.361$   \\
$\textbf{p(F)=0.000}$             & -                                   &   -                                  & 
$\textbf{p(F)=0.000}$             & $\textbf{p(F)=0.000}$               & $\textbf{p(F)=0.000}$         \\ \hline
$P=2.440\pm0.014$~d  &   $P=2.466\pm0.013$~d  &   $P=2.540\pm0.040$~d      &   $P=2.433\pm0.028$~d  &   $P=2.449\pm0.029$~d  & $P=2.454\pm0.041$~d  \\
$F=6.681$  &   $F=6.332$    &   $F=8.576$        &   $F=16.248$   &   $F=15.572$       & $F=19.553$ \\
$\textbf{p(F)=0.000}$             & $\textbf{p(F)=0.001}$               & $\textbf{p(F)=0.000}$               &
$\textbf{p(F)=0.000}$             & $\textbf{p(F)=0.000}$               & $\textbf{p(F)=0.000}$       \\ \hline
$P=1.050\pm0.028$~d  &   $P=1.070\pm0.018$~d  &   $P=0.980\pm0.083$~d     &   $P=1.039\pm0.054$~d    &   $P=0.946\pm0.061$~d     &   $P=0.579\pm0.053$~d   \\
$F=5.562$   &   $F=5.695$   &   $F=1.709$      &   $F=9.339$       &   $F=9.280$       &   $F=11.431$ \\
$\textbf{p(F)=0.003}$             & $\textbf{p(F)=0.003}$               & $p(F)=0.181$             &
$\textbf{p(F)=0.000}$             & $\textbf{p(F)=0.000}$               & $\textbf{p(F)=0.000}$   \\ \hline
$P=0.350\pm0.014$~d  & -             &   $P=3.890\pm0.436$~d     &   $P=0.346\pm0.032$~d     &  $P=0.659\pm0.103$~d    &   $P=0.352\pm0.019$~d   \\ 
$F=6.944$   &   -           &   $F=11.259$      &   $F=6.716$       &  $F=6.696$      &   $F=6.847$  \\  
$\textbf{p(F)=0.000}$       & -                 &  $\textbf{p(F)=0.000}$                &   \textbf{p(F)=0.001}              &   $\textbf{p(F)=0.001}$               &   $\textbf{p(F)=0.001}$  \\ \hline \hline 
\multicolumn{6}{c}{SMC}\\ \hline
$P=2.5$~d   &   $P=2.5$~d   &   $P=2.5$~d       &   $P=2.5$~d   &   $P=2.5$~d       &   $P=2.5$~d      \\
$F=9.656$   &   $F=8.340$   &   $F=4.816$       &   $F=4.660$   &   $F=7.462$       &   $F=6.257$ \\ 
$\textbf{p(F)=0.000}$             & $\textbf{p(F)=0.000}$               & $\textbf{p(F)=0.008}$             &
$\textbf{p(F)=0.009}$             & $\textbf{p(F)=0.000}$               & $\textbf{p(F)=0.002}$          \\ \hline
$P=2.494\pm0.018$~d &   $P=2.506\pm0.037$~d &   $P=2.454\pm0.106$~d      &   $P=2.296\pm0.161$~d  &   $P=2.338\pm0.180$~d     &   $P=2.344\pm0.229$~d    \\
$F=9.656$   &   $F=8.293$   &   $F=5.088$        &   $F=6.085$    &   $F=8.116$       &   $F=6.806$ \\
$\textbf{p(F)=0.000}$             & $\textbf{p(F)=0.000}$               & $\textbf{p(F)=0.003}$             &
$\textbf{p(F)=0.003}$             & $\textbf{p(F)=0.000}$               & $\textbf{p(F)=0.001}$          \\ \hline
$P=1.023\pm0.017$~d &   $P=1.013\pm0.093$~d &   $P=1.078\pm0.073$~d     &   $P=1.013\pm0.073$~d    &   $P=0.961\pm0.080$~d     &   $P=0.926\pm0.080$~d    \\ 
$F=14.076$  &   $F=3.240$   &   $F=3.493$       &   $F=7.164$      &   $F=7.084$       &   $F=4.590$   \\
$\textbf{p(F)=0.000}$   &  $\textbf{p(F)=0.039}$   &    $\textbf{p(F)=0.028}$   &   $\textbf{p(F)=0.000}$   & $\textbf{p(F)=0.000}$   &   $\textbf{p(F)=0.010}$          \\ \hline \hline
\end{tabular}
\label{tab:fvalue_pc_lmcsmc}
\end{center}
\end{table*}

\begin{table*}
\begin{center}
\caption{Results of the $F$-test for break-points in the PC relations using piecewise regression from the observed data at mean light for the LMC and SMC FU Cepheids. The bold-faced entries indicate the significance of the break-points in the PC relations. The first ($P=2.5$ d) row indicate the results of the $F$- test for the corresponding break-points as reported in the literature. The remaining rows represent the $F$-test results for break-points obtained using piecewise regression. \textbf{The uncertainties represent standard errors based on $95\%$ confidence interval obtained from piecewise regression.}}
\begin{tabular}{c c c c c c c} \\ \hline \hline
$V-I$    &    $V-G$   & $V-G_{RP}$  & $V-Y$ & $V-J$   & $V-Ks$   \\ \hline
\multicolumn{6}{c}{LMC}\\ \hline
$P=2.5$~d   &   $P=2.5$~d   &   $P=2.5$~d     &   $P=2.5$~d     &   $P=2.5$~d       & $P=2.5$~d \\
$F=2.551$  &   $F=1.088$   &   $F=0.621$     &   $F=0.579$    &   $F=0.174$      & $F=11.692$ \\
$p(F)=0.078$             & $p(F)=0.336$               &   $p(F)=0.537$             &
$p(F)=0.560$             & $p(F)=0.840$               &   $p(F)=0.320$ \\ \hline
$P=2.523\pm0.013$~d &   $P=2.483\pm0.092$~d  &   $P=2.483\pm0.127$~d      &   $P=2.535\pm0.031$~d &   $P=2.494\pm0.032$~d      & $P=2.523\pm0.032$~d  \\
$F=2.790$   &   $F=0.915$    &   $F=1.840$        &   $F=0.697$   &   $F=0.103$        & $F=1.743$ \\
$p(F)=0.061$    &   $p(F)=0.400$    & $p(F)=0.159$  &   $p(F)=0.498$    & $p(F)=0.902$  & $p(F)=0.175$       \\ \hline
$P=4.027\pm0.049$~d &   $P=4.036\pm0.324$~d &   $P=3.907\pm0.255$~d     &   $P=4.073\pm0.105$~d    &   $P=3.962\pm0.116$~d     &   $P=4.045\pm0.120$~d   \\
$F=0.957$   &   $F=0.608$   &   $F=0.263$       &   $F=0.209$      &   $F=0.274$       &   $F=0.452$ \\
$p(F)=0.384$    &   $p(F)=0.544$    &   $p(F)=0.792$    &   $p(F)=0.811$    & $p(F)=0.760$  & $p(F)=0.636$   \\ \hline
$P=6.591\pm0.965$~d & $P=8.413\pm1.506$     &   -     &   -     &  -    &   -   \\ 
$F=3.035$   & $F=1.294$     &   -     &   -     &  -    &   -  \\  
$\textbf{p(F)=0.048}$       &   $p(F)=0.274$     &   -     &   -     &   -  &   -  \\ \hline \hline 
\multicolumn{6}{c}{SMC}\\ \hline
$P=2.5$~d   &   $P=2.5$~d   &   $P=2.5$~d       &   $P=2.5$~d   &   $P=2.5$~d       &   $P=2.5$~d      \\
$F=6.146$   &   $F=5.964$   &   $F=4.816$       &   $F=4.660$   &   $F=7.462$       &   $F=6.257$ \\ 
$\textbf{p(F)=0.000}$             & $\textbf{p(F)=0.000}$               & $\textbf{p(F)=0.008}$             &
$\textbf{p(F)=0.009}$             & $\textbf{p(F)=0.000}$               & $\textbf{p(F)=0.002}$          \\ \hline
$P=2.654\pm0.098$~d &   $P=2.517\pm0.046$~d &   $P=2.409\pm0.022$~d      &   $P=2.322\pm0.048$~d  &   $P=2.371\pm0.052$~d     &   $P=2.275\pm0.050$~d    \\
$F=6.643$   &   $F=5.754$   &   $F=16.611$       &   $F=3.309$    &   $F=7.300$       &   $F=6.957$ \\
$\textbf{p(F)=0.001}$             & $\textbf{p(F)=0.003}$               & $\textbf{p(F)=0.000}$             &
$\textbf{p(F)=0.036}$             & $\textbf{p(F)=0.000}$               & $\textbf{p(F)=0.000}$          \\ \hline
$P=1.023\pm0.090$~d &   $P=2.128\pm0.070$~d &   $P=4.045\pm0.169$~d     &   $P=4.315\pm0.498$~d    &   $P=3.815\pm0.405$~d     &   $P=7.550\pm0.803$~d    \\ 
$F=14.076$  &   $F=5.403$   &   $F=2.144$       &   $F=1.846$      &   $F=3.441$       &   $F=1.649$   \\
$\textbf{p(F)=0.000}$   &  $\textbf{p(F)=0.003}$   &    $p(F)=0.121$   &   $p(F)=0.158$   & $\textbf{p(F)=0.032}$   &   $p(F)=0.192$          \\ \hline 
-   &   $P=3.999\pm0.158$   &   -   &   -   &   -   &   -      \\
-   &   $F=3.496$   &   -   &   -   &   -   &   -      \\
-   &   $\textbf{p(F)=0.031}$    &   -   &   -   &   -   &   -      \\ \hline \hline
\end{tabular}
\label{tab:fvalue_pc_fu_lmcsmc}
\end{center}
\end{table*}

\begin{table*}
\begin{center}
\caption{Same as Table \ref{tab:fvalue_pc_lmcsmc} but for PL relations.} 
\scalebox{0.95}{
\begin{tabular}{c c c c c c c c} \\ \hline \hline
$V$ & $I$   &   $G$ & $G_{RP}$  &  $Y$   & $J$   & $Ks$   \\ \hline
\multicolumn{7}{c}{LMC}\\ \hline
$P=2.5$~d   &   $P=2.5$~d   &   $P=2.5$~d   &   $P=2.5$~d       &   $P=2.5$~d   &   $P=2.5$~d   &   $P=2.5$~d   \\
$F=20.787$  &   $F=19.346$  &   $F=15.986$  &   $F=13.781$      &   $F=22.502$  &   $F=30.690$  &   $F=29.291$  \\
$\textbf{p(F)=0.000}$           &   $\textbf{p(F)=0.000}$          &   $\textbf{p(F)=0.000}$        &
$\textbf{p(F)=0.000}$           &   $\textbf{p(F)=0.000}$          &   $\textbf{p(F)=0.000}$        & 
$\textbf{p(F)=0.000}$   \\ \hline
$P=0.58$~d  &   $P=0.58$~d  &  -            &   -               &   $P=0.58$~d  &   $P=0.58$~d  &  $P=0.58$~d  \\
$F=9.417$   &   $F=4.918$   &  -            &   -               &   $F=7.123$   &   $F=7.123$   &  $F=9.516$   \\ 
$\textbf{p(F)=0.000}$       &  $\textbf{p(F)=0.007}$                &   -                           &
-                           &   $\textbf{p(F)=0.000}$               &   $\textbf{p(F)=0.000}$       &
$\textbf{p(F)=0.000}$    \\ \hline
$P=2.390\pm0.055$~d  &   $P=2.460\pm0.037$~d  &   $P=2.506\pm0.074$~d &  $P=2.443\pm0.086$~d      &   $P=2.506\pm0.042$~d  &   $P=2.460\pm0.039$~d  &   $P=2.488\pm0.030$~d  \\
$F=21.704$  &   $F=19.444$  &   $F=15.912$  &  $F=15.340$       &   $F=22.564$  &   $F=30.294$   &   $F=24.442$ \\
$\textbf{p(F)=0.000}$           & $\textbf{p(F)=0.000}$                    & $\textbf{p(F)=0.000}$            &
$\textbf{p(F)=0.000}$           & $\textbf{p(F)=0.000}$             & $\textbf{p(F)=0.000}$            &
$\textbf{p(F)=0.000}$   \\ \hline
$P=0.612\pm0.189$~d &   $P=0.410\pm0.028$~d  &   $P=3.499\pm0.457$~d &   -               &   $P=0.594\pm0.039$~d   &   $P=0.582\pm0.035$~d & $P=0.433\pm0.052$~d \\
$F=12.313$   &   $F=3.562$   &   $F=13.616$  &   -               &   $F=10.829$    &   $F=12.182$  & $F=3.635$  \\
$\mathbf{p(F)=0.047}$ &   $\textbf{p(F)=0.027}$    &   $\textbf{p(F)=0.000}$ &   -   & $\textbf{p(F)=0.000}$ &   $\textbf{p(F)=0.000}$ & $\textbf{p(F)=0.026}$    \\ \hline
-           &   $P=0.368\pm0.027$~d & -             &   -               &   $P=3.710\pm0.291$~d  &   $P=0.406\pm0.034$~d &   $P=0.629\pm0.047$~d \\
-           &   $F=3.798$   & -             &   -               &   $F=10.426$  &   $F=6.295$   &   $F=15.470$   \\ -           &   $\textbf{p(F)=0.022}$              &  -            &   -               &   \textbf{p(F)=0.000}  &   \textbf{p(F)=0.000}  & \textbf{p(F)=0.000} \\ \hline
-           &   -           &   -           &   -               &   $P=0.407\pm0.035$~d &   -   &   -       \\
-           &   -           &   -           &   -               &   $F=2.538$   &   -   &   - \\  
-           &   -           &   -           &   -               &   $p(F)=0.079$    &   -   &   - \\ \hline \hline          

\multicolumn{7}{c}{SMC}\\ \hline
$P=2.5$~d   &   $P=2.5$~d    &   $P=2.5$~d   &   $P=2.5$~d       &   $P=2.5$~d     & $P=2.5$~d     &   $P=2.5$~d \\ 
$F=8.396$   &   $F=7.184$    &   $F=5.931$   &   $F=4.973$       &   $F=7.619$     &   $F=5.995$   &   $F=5.129$ \\
$\textbf{p(F)=0.000}$           &   $\textbf{p(F)=0.000}$           &   $\textbf{p(F)=0.003}$       &
$\textbf{p(F)=0.006}$           &   $\textbf{p(F)=0.000}$           &   $\textbf{p(F)=0.002}$       &   
$\textbf{p(F)=0.002}$   \\ \hline
$P=2.576\pm0.108$~d &   $P=2.471\pm0.081$~d  &   $P=2.570\pm0.128$~d &   $P=2.624\pm0.195$~d &   $P=2.750\pm0.120$~d  &   $P=2.630\pm0.113$~d  & $P=2.618\pm0.115$~d \\
$F=8.788$   &   $F=7.167$    &   $F=6.038$   &   $F=5.027$   &   $F=7.124$   &   $F=5.878$   &  $F=4.818$ \\ 
$\textbf{p(F)=0.000}$        &  $\textbf{p(F)=0.000}$            & $\textbf{p(F)=0.000}$         &
$\textbf{p(F)=0.002}$        & $\textbf{p(F)=0.006}$             & $\textbf{p(F)=0.000}$         &
$\textbf{p(F)=0.006}$   \\ \hline
$P=0.990\pm0.050$~d &   $P=1.000\pm0.045$~d &   $P=1.000\pm0.066$~d  &   $P=0.954\pm0.081$~d &   $P=0.830\pm0.095$~d  &   $P=1.086\pm0.148$~d &  $P=1.074\pm0.150$~d \\
$F=4.736$   &   $F=2.874$    &   $F=1.622$   &   $F=1.930$   &   $F=2.281$   &   $F=0.524$   &  $F=1.303$\\ 
$p(F)=0.062$   &   $p(F)=0.056$   &   $p(F)=0.145$         & $p(F)=0.197$   & 
$p(F)=0.102$                & $p(F)=0.592$         & $p(F)=0.288$   \\ \hline
\hline
\end{tabular}}
\label{tab:fvalue_pl_lmcsmc}
\end{center}
\end{table*}

\begin{table*}
\begin{center}
\caption{Same as Table \ref{tab:fvalue_pc_fu_lmcsmc} but for PL relations.} 
\scalebox{0.95}{
\begin{tabular}{c c c c c c c c} \\ \hline \hline
$V$ & $I$   &   $G$ & $G_{RP}$  &  $Y$   & $J$   & $Ks$   \\ \hline
\multicolumn{7}{c}{LMC}\\ \hline
$P=2.5$~d   &   $P=2.5$~d   &   $P=2.5$~d   &   $P=2.5$~d       &   $P=2.5$~d   &   $P=2.5$~d   &   $P=2.5$~d   \\
$F=2.272$  &   $F=2.503$  &   $F=1.000$  &   $F=1.170$      &   $F=2.178$  &   $F=1.753$  &   $F=1.600$  \\
$p(F)=0.103$           &   $p(F)=0.082$          &   $p(F)=0.367$        &
$p(F)=0.310$           &   $p(F)=0.113$          &   $p(F)=0.173$        & 
$p(F)=0.202$   \\ \hline
$P=2.540\pm0.044$~d &   $P=2.612\pm0.043$~d &   $P=2.642\pm0.087$~d &  $P=2.471\pm0.115$~d  &   $P=2.482\pm0.122$~d &   $P=2.423\pm0.045$~d &   $P=2.529\pm0.022$~d  \\
$F=2.145$   &   $F=3.271$   &   $F=1.702$   &  $F=2.726$    &   $F=1.242$   &   $F=1.220$   &   $F=2.239$ \\
$p(F)=0.117$                & $\textbf{p(F)=0.038}$     & $p(F)=0.183$      &   $p(F)=0.065$    & $p(F)=0.337$
& $p(F)=0.295$  &   $p(F)=0.106$   \\ \hline
$P=4.092\pm0.170$~d &   $P=4.083\pm0.118$~d &   $P=4.092\pm0.395$~d &   $P=4.365\pm0.566$~d     &   $P=1.823\pm0.113$~d &   -    & $P=4.045\pm0.076$~d \\
$F=1.757$   &   $F=2.452$   &   $F=4.006$   &   $F=1.354$       &   $F=8.537$   &   -    & $F=1.087$  \\
$p(F)=0.172$    &   $p(F)=0.086$    &   \textbf{p(F)=0.018}         &   $p(F)=0.258$           & $\textbf{p(F)=0.000}$        
            &   -               & $p(F)=0.206$    \\ \hline
-           &   $P=1.798\pm0.122$~d & -             &   -               &   -             &   - &   - \\
-           &   $F=2.270$   & -             &   -               &   -             &   - &   -\\
-           &   $p(F)=0.103$                &   -               &   -             &   - &  -   \\ \hline \hline     
\multicolumn{7}{c}{SMC}\\ \hline
$P=2.5$~d   &   $P=2.5$~d    &   $P=2.5$~d   &   $P=2.5$~d  &   $P=2.5$~d     & $P=2.5$~d     &   $P=2.5$~d \\ 
$F=21.070$  &   $F=23.446$   &   $F=20.640$  &   $F=22.347$  &   $F=17.492$    & $F=17.759$   &   $F=22.347$ \\
$\textbf{p(F)=0.000}$           &   $\textbf{p(F)=0.000}$           &   $\textbf{p(F)=0.003}$       &
$\textbf{p(F)=0.006}$           &   $\textbf{p(F)=0.000}$           &   $\textbf{p(F)=0.002}$       &   
$\textbf{p(F)=0.002}$   \\ \hline
$P=2.249\pm0.052$~d &   $P=2.421\pm0.303$~d  &  $P=2.381\pm0.058$~d &   $P=2.576\pm0.114$~d &   $P=2.506\pm0.064$~d  &   $P=2.654\pm0.055$~d  &   $P=2.404\pm0.063$~d  \\
$F=19.752$  &   $F=24.805$   &  $F=21.872$  &   $F=15.998$   &   $F=17.715$   &   $F=15.395$   &   $F=21.439$ \\ 
$\textbf{p(F)=0.000}$        &  $\textbf{p(F)=0.000}$       & $\textbf{p(F)=0.000}$ &   $\textbf{p(F)=0.000}$       & $\textbf{p(F)=0.000}$      & $\textbf{p(F)=0.000}$        & $\textbf{p(F)=0.000}$   \\ \hline
$P=5.712\pm0.044$~d &   $P=2.679\pm0.398$~d   & $P=3.681\pm0.239$~d &   $P=2.055\pm0.070$~d &   $P=3.013\pm0.147$~d  &   $P=2.910\pm0.042$~d &   $P=4.852\pm0.341$~d \\
$F=7.890$   &   $F=22.813$    & $F=0.928$   &   $F=22.320$  &   $F=13.104$   &   $F=14.689$  &   $F=12.456$\\ 
$\textbf{p(F)=0.000}$   &   $\textbf{p(F)=0.000}$   &   $p(F)=0.395$         & $\textbf{p(F)=0.000}$   & 
$\textbf{p(F)=0.000}$                & $\textbf{p(F)=0.000}$         & $\textbf{p(F)=0.000}$   \\ \hline
-   &   -   &   $P=1.566\pm0.046$~d   &   -   &   -   &   -   &   $P=6.025\pm0.790$~d \\
-   &   -   &   $F=24.249$   &   -   &   -   &   -   &   $F=10.680$ \\
-   &   -   &   \textbf{p(F)=0.000}   &   -   &   -   &   -   &   \textbf{p(F)=0.000} \\ \hline \hline
\end{tabular}}
\label{tab:fvalue_pl_fu_lmcsmc}
\end{center}
\end{table*}

\section{$t$ test for the PL relations of FO Cepheids in the LMC/SMC}
\label{sec:t_test_pl_appen}

The results of $t$ test of the PL relation of FO Cepheids obtained using the models and observations are summarised in Table~\ref{tab:pl_models_lmc} for the LMC and \ref{tab:pl_models_smc} for the SMC.

\begin{table*}
\begin{center}
\caption{Comparison of the PL relations for the FO Cepheids in the LMC obtained using the observations and models. $|T|, p(t)$ represents the observed value and the probability of the $t$statistics. Bold-faced entries indicate the null hypothesis (equal slopes) can be rejected.}
\scalebox{0.95}{
\begin{tabular}{c c c c c c c c c c c c c c} \\ \hline \hline
Band&    Source   & $a_{\rm all}$ & $b_{\rm all}$ & $\sigma$ & N & Reference & Theoretical/Empirical & \multicolumn{2}{c}{$|T|,p(t) w.r.t$} \\
    &             &           &           &   &           &           &     &       & Set B  & Set D \\ \hline 
$V$   & Set B   & $-3.189\pm0.029$  &  $-1.879\pm0.015$ & $0.109$   &   $263$  & TW    & Theoretical
                & ... & ...  \\  
$V$   & Set D   & $-2.921\pm0.033$  &  $-1.882\pm0.018$ & $0.119$   &   $265$ & TW    & Theoretical
                & \textbf{(17.018, 0.000)} & ...  \\  
$V$   & OGLE-IV & $-3.291\pm0.025$  & $16.739\pm0.008$  & $0.175$   & $1355$                    & TW    & Empirical 
                & \textbf{(6.940, 0.020)} & \textbf{(17.176, 0.000)} \\
$I$   & Set B   & $-3.350\pm0.023$  &  $-2.381\pm0.012$ & $0.093$   & $263$  & TW    & Theoretical
                & ... & ...  \\  
$I$   & Set D   & $-3.148\pm0.027$  &  $-2.404\pm0.013$ & $0.096$   & $265$ & TW    & Theoretical	
                & \textbf{(14.283, 0.000)} & ...  \\  
$I$   & OGLE-IV & $-3.349\pm0.018$  & $16.203\pm0.006$  & $0.126$   & $1355$                    & TW    & Empirical
                & (1.054, 0.069)    & \textbf{(9.287, 0.000)}  \\
$G$   & Set B   &   $-3.249\pm0.027$  &  $-2.017\pm0.014$ & $0.105$   & $263$  & TW    & Theoretical
                & ... & ...  \\  
$G$   & Set D   &  $-2.999\pm0.032$  &  $-2.031\pm0.017$  & $0.112$   & $265$ & TW    & Theoretical	
                & \textbf{(14.555, 0.000)} & ...  \\  
$G$   & Gaia-DR3    & $-3.281\pm0.028$ & $16.656\pm0.010$  & $0.164$   & $795$                      & TW    & Empirical
                    &  \textbf{(4.314, 0.000)} & \textbf{(18.203, 0.000)} \\
$G_{RP}$   & Set B  &  $-3.335\pm0.032$  &  $-2.353\pm0.017$    & $0.143$  & $263$  & TW    & Theoretical   
                    & ... & ...  \\  
$G_{RP}$   & Set D  &  $-3.129\pm0.028$  &  $-2.403\pm0.016$    & $0.099$  & $265$ & TW    & Theoretical
                    & \textbf{(13.297, 0.000)} & ...  \\  
$G_{RP}$   & Gaia-DR3   & $-3.369\pm0.023$    & $16.256\pm0.008$  & $0.137$   
                        & $988$ & TW & Empirical
                        & (1.528, 0.052) & \textbf{(15.029,0.000)} \\
$J$   & Set B   &  $-3.459\pm0.021$  &  $-2.721\pm0.011$    & $0.086$   & $263$ & TW    & Theoretical
                & ... & ...  \\  
$J$   & Set D   &  $-3.288\pm0.024$  &  $-2.757\pm0.014$    & $0.085$   & $265$ & TW    & Theoretical	
                & \textbf{(14.717, 0.000)} & ...  \\  
$J$ & VMC       & $-3.394\pm0.015$  & $15.824\pm0.005$  & $0.118$   & $1291$                    &  TW   & Empirical
                & \textbf{(4.422, 0.000)} & \textbf{(5.657, 0.000)} \\ 
$Ks$  & Set B   &  $-3.540\pm0.020$  &  $-2.946\pm0.010$    & $0.081$   & $263$  & TW    & Theoretical
                & ... & ...  \\  
$Ks$  & Set D   &  $-3.404\pm0.022$  &  $-2.993\pm0.013$    & $0.077$   & $265$ & TW    & Theoretical
                & \textbf{(14.947, 0.000)} & ...  \\  
$Ks$  & VMC     & $-3.452\pm0.011$  & $15.543\pm0.004$  & $0.087$   & $1291$
                & TW    & Empirical  & \textbf{(5.328, 0.000)} & (2.519, 0.293)\\ \hline \hline 
\end{tabular}
\label{tab:pl_models_lmc}}
\end{center}
\end{table*}

\begin{table*}
\begin{center}
\caption{Same as Table \ref{tab:pl_models_smc} but for SMC.}
\scalebox{0.95}{
\begin{tabular}{c c c c c c c c c c c c c c} \\ \hline \hline
Band&    Source   & $a_{\rm all}$ & $b_{\rm all}$ & $\sigma$ & N & Reference & Theoretical/Empirical & \multicolumn{2}{c}{$|T|,p(t) w.r.t$} \\
    &             &           &           &   &           &                       & Set B  & Set D \\ \hline 
$V$   & Set B   & $-3.108\pm0.043$  & $-1.968\pm0.017$  & $0.190$   & $224$     & TW    & Theoretical	
                & ... & ...  \\  
$V$   & Set D   & $-3.029\pm0.034$  & $-1.860\pm0.015$ & $0.153$   & $233$     & TW    & Theoretical	
                & \textbf{(3.000, 0.140)}   &  ...  \\  
$V$   & OGLE-IV & $-3.166\pm0.033$  & $17.180\pm0.008$  & $0.266$   & $1532$ 
                & TW    & Empirical & \textbf{(2.103, 0.000)}    & \textbf{(16.737, 0.000)} \\
$I$   & Set B   & $-3.445\pm0.028$  & $-2.388\pm0.012$  & $0.174$   & $224$    & TW    & Theoretical	
                & ... & ...  \\  
$I$   & Set D   & $-3.190\pm0.030$  & $-2.388\pm0.013$ & $0.140$   & $233$     & TW    & Theoretical	
                & \textbf{(23.676, 0.000)}    & ... \\  
$I$   & OGLE-IV & $-3.321\pm0.027$  & $16.696\pm0.006$  & $0.219$   & $1532$ 
                & TW    & Empirical & \textbf{(16.720, 0.000)}    & \textbf{(10.017, 0.000)}\\
$G$   & Set B   & $-3.165\pm0.041$  & $-2.102\pm0.015$  & $0.185$   & $224$     & TW    & Theoretical	
                & ... & ...  \\  
$G$   & Set D   & $-3.089\pm0.032$  & $-2.008\pm0.014$  & $0.148$   & $233$     & TW    & Theoretical	
                & \textbf{(2.965, 0.001)}    & ...  \\  
$G$   & Gaia-DR3    & $-3.196\pm0.036$  & $17.119\pm0.009$  &  $0.251$  & $1182$    & TW    & Empirical 
                    & \textbf{(1.579, 0.057)}    & \textbf{(6.487, 0.000)} \\
$G_{RP}$   & Set B  & $-3.258\pm0.038$  & $-2.446\pm0.025$  & $0.176$   & $224$     & TW    & Theoretical	
                    & ... & ...  \\  
$G_{RP}$   & Set D  & $-3.184\pm0.030$  & $-2.380\pm0.013$  & $0.140$   & $233$     & TW    & Theoretical	
                    & \textbf{(3.172, 0.000)}    & ...  \\  
$G_{RP}$   & Gaia-DR3   & $-3.297\pm0.033$  & $16.745\pm0.008$  & $0.227$   & $1162$    & TW    & Empirical 
                        & \textbf{(2.069, 0.019)}    & \textbf{(8.219, 0.000)}\\
$J$   & Set B   & $-3.368\pm0.035$  & $-2.782\pm0.015$  & $0.166$   & $224$    & TW    & Theoretical	
                & ... & ...  \\  
$J$   & Set D   & $-3.299\pm0.027$  & $-2.745\pm0.012$  & $0.133$   & $233$     & TW    & Theoretical	
                & \textbf{(3.098, 0.000)}    & ...  \\ 
$J$   & VMC     & $-3.412\pm0.025$  & $16.350\pm0.006$  & $0.195$   & $1425$    & TW    & Emperical
                & \textbf{(1.796, 0.036)}    & \textbf{(11.080, 0.000)} \\
$Ks$   & Set B  & $-3.448\pm0.033$  & $-3.003\pm0.014$  & $0.159$   & $224$     & TW    & Theoretical	
                & ... & ...  \\  
$Ks$   & Set D  & $-3.384\pm0.026$  & $-2.988\pm0.011$  & $0.128$     & $233$     & TW    & Theoretical	
                & \textbf{(3.149,0.000)} & ...  \\  
$Ks$   & VMC    & $-3.539\pm0.022$  & $16.123\pm0.005$  & $0.173$   & $1425$    & TW    & Empirical 
                & \textbf{(3.699, 0.000)}    & \textbf{(11.186, 0.000)} \\ \hline  \hline
\end{tabular}
\label{tab:pl_models_smc}}                                
\end{center}
\end{table*}

\bsp	
\label{lastpage}
\end{document}